\documentclass[a4paper,11pt]{article}
\pdfoutput=1 % if your are submitting a pdflatex (i.e. if you have
\usepackage{jheppub}

\usepackage{graphicx, epstopdf}

\usepackage[utf8]{inputenc}

\usepackage{amsmath, amsthm, amsfonts, amssymb}
\usepackage{mathtools, mathabx, bm, bbm, scalerel, xfrac}
\usepackage{empheq, physics, slashed, cancel}
\usepackage[usenames,dvipsnames]{xcolor}

\usepackage{hyperref}
\usepackage{color}

\allowdisplaybreaks

\title{Subleading Effects in Soft-Gluon Emission at One-Loop in Massless QCD}

\author{Micha\l{} Czakon,}
\author{Felix Eschment and}
\author{Tom Schellenberger}
\affiliation{Institut f\"ur Theoretische Teilchenphysik und Kosmologie, RWTH Aachen University,\\ D-52056 Aachen, Germany}
\emailAdd{mczakon@physik.rwth-aachen.de}
\emailAdd{felix.eschment@rwth-aachen.de}
\emailAdd{tom.schellenberger@rwth-aachen.de}

\abstract{We elucidate the structure of the next-to-leading-power soft-gluon expansion of arbitrary one-loop massless-QCD amplitudes. The expansion is given in terms of universal colour-, spin- and flavour-dependent operators acting on process-dependent gauge-invariant amplitudes. The result is proven using the method of expansion-by-regions and tested numerically on non-trivial processes with up to six partons. In principle, collinear-region contributions are expressed in terms of convolutions of universal jet operators and process-dependent amplitudes with two collinear partons. However, we evaluate these convolutions exactly for arbitrary processes. This is achieved by deriving an expression for the next-to-leading power expansion of tree-level amplitudes in the collinear limit, which is a novel result as well. Compared to previous studies, our analysis, besides being more general, yields simpler formulae that avoid derivatives of process-dependent amplitudes in the collinear limit.}
\preprint{P3H-23-041, TTK-23-17}

\begin{document}
\maketitle
\flushbottom

\section{Introduction}

Soft radiation is an important topic in the context of gauge theories. In the abelian case of QED, soft photons are physical and complicate the definition of the scattering operator. In the non-abelian case, in particular in QCD, gauge bosons are not physical in the confining phase, and the presence of a mass gap protects from soft singularities. However, in the context of factorisation, which allows to obtain cross sections as a convolution of a non-perturbative contribution and a contribution that involves massless partons, the problem appears again. In either case, abelian and non-abelian, it is necessary to have a complete description of the leading singular soft asymptotics in order to obtain meaningful theoretical predictions for scattering and decay processes. This problem has been studied since the early days of Quantum Field Theory and is nowadays textbook material. While the subleading behaviour of scattering amplitudes in the soft limit is not necessary to obtain finite cross sections, it is nevertheless of interest due to the ever increasing precision of measurements at lepton and hadron colliders. First attempts at a general description in QED date back to the seminal works of Low \cite{Low:1958sn}, Burnett and Kroll \cite{Burnett:1967km}. Later, it was understood by Del Duca \cite{DelDuca:1990gz} that the description cannot be complete beyond tree-level without taking into account collinear virtual states.

Recently, there has been a surge of interest in next-to-leading power (subleading) soft phenomena within resummation formalisms based on \textit{Soft-Collinear Effective Theory} (SCET) \cite{Larkoski:2014bxa,Beneke:2019oqx,Liu:2021mac} (see also related studies in gravity \cite{Beneke:2021umj,Beneke:2021aip,Beneke:2022pue}) and diagrammatic approaches to QCD \cite{Bonocore:2015esa,Bonocore:2016awd}. The main goal of the studies was the inclusion of subleading effects in the description of simple processes with a minimal number of partons, for example the Drell-Yan process. Even in this case, there were surprises and some assumptions on the structure of the soft expansion turned out to be wrong. For instance, the analysis of Ref.~\cite{DelDuca:1990gz} that introduced collinear radiation into the picture, was shown to be incomplete.

A different motivation for studying subleading soft effects in QED with massive fermions guided Refs.~\cite{Engel:2021ccn, Engel:2023ifn}. Here, the idea was to use soft approximations of squared matrix elements to obtain numerically stable predictions for lepton scattering with account of soft photons and light leptons.

Our goal in the present publication is to understand the structure of the next-to-leading-power soft expansion at the one-loop level in QCD. On the one hand, the general expression that we derive allows to put resummation formalisms for multi-parton processes on a firm footing. On other hand, this expression can be used to improve the numerical stability of matrix elements in software implementations.

In our analysis, we stress not only the importance of the Ward identity for the soft gluon -- as did the pioneers -- but also of gauge-invariance of the occurring amplitudes. This leads to astonishingly simple expressions for the building blocks of the expansion: soft and jet operators. The cancellations that we observe remove contributions that are expected to be present based on pure power-counting arguments, for example transverse-momentum derivatives of amplitudes in the collinear limit, see Refs.~\cite{Gervais:2017yxv, Laenen:2020nrt}. Furthermore, we put special emphasis on a deep understanding of the collinear asymptotics. As a side effect, we obtain a novel formula for the next-to-leading power expansion of tree-level amplitudes in the collinear limit.

The publication is organised as follows. In the next section we define the main concepts and recall the colour/spin-space formalism that proves to be very useful in the present context. We define spin-space operators that encapsulate all spin effects at next-to-leading power. We also take great care to define the kinematics of the soft limit to the level of detail required in a numerical application. In Section~\ref{sec:LBK}, we reproduce the Low-Burnett-Kroll result for QCD, and summarise its features that have been understood in previous studies. We use, nevertheless, our original notation that will prove its power at the one-loop level. In Section~\ref{sec:OneLoopLBK}, we state our main result, present a complete proof, and describe numerical tests. Finally, in Section~\ref{sec:CollinearNLP}, we state our result for the next-to-leading-power collinear asymptotics. An outlook section closes the text and discusses some obvious further directions of research.

\section{Definitions}

\subsection{Processes and amplitudes}

Consider the process:
\begin{equation} \label{eq:process}
    0 \to a_1(p_1 + \delta_1, \sigma_1, c_1) + \dots + a_n(p_n + \delta_n, \sigma_n, c_n) + g(q, \sigma_{n+1}, c_{n+1}) \; , \qquad a_i \in \{q, \bar{q}, g\} \; .
\end{equation}
The momenta $p_i + \delta_i$ of the {\it hard partons} are defined as outgoing, and may thus have negative energy components if the respective parton is actually incoming in the physical process under consideration. The {\it soft gluon} with momentum $q$ is outgoing, $q^0 > 0$. The momenta are assumed on-shell:
\begin{equation} \label{eq:onshellness}
    p_i^2 = (p_i + \delta_i)^2 = m_i^2 \; , \qquad q^2 = 0 \; ,
\end{equation}
where $m_i$ is the mass of parton $i$. The {\it momentum shifts}, $\delta_i$, are introduced to ensure that the two sets of momenta, $\{\{p_i+\delta_i\}_{i=1}^n,q\}$ and $\{p_i\}_{i=1}^n$, both satisfy momentum conservation by requiring:
\begin{equation} \label{eq:momentumconservation}
    \sum_i p_i = 0 \; , \qquad \sum_i \delta_i + q = 0 \; .
\end{equation}
Notice that Eqs.~\eqref{eq:onshellness} and \eqref{eq:momentumconservation} are more restrictive than necessary for a physical process. The additional constraints are used to define the soft limit. Contrary to the hard momenta, $p_i$, every component of the momentum shifts and every component of the soft-gluon momentum is assumed to be of the order of the {\it soft-expansion parameter} $\lambda$:
\begin{equation} \label{eq:order}
    p_i^\mu = \order{1} = \order{\lambda^0} \gg \lambda \; , \qquad \delta_i^\mu = \order{\lambda} \; , \qquad q^\mu  = \order{\lambda} \; .
\end{equation}
Finally, $p_i$ and $q$ are assumed well separated in angular distance. It follows from Eqs.~\eqref{eq:onshellness} and \eqref{eq:order} that $p_i$ is orthogonal to $\delta_i$ in first approximation:
\begin{equation} \label{eq:orthogonality}
    p_i \cdot \delta_i = \order{\lambda^2} \; .
\end{equation}
The polarisation and colour state of each parton is denoted by $\sigma_i$ and $c_i$ respectively. The polarisation of massive partons may be defined as rest-frame spin, whereas that of massless partons corresponds to helicity.

The results of this publication are equally {\it valid in the case of quarks of different flavours as well as in the presence of colour-neutral particles}, as long as flavour and colour summations have been appropriately adapted.

A scattering amplitude, $M_{fi}$, is defined through the decomposition of the scattering matrix $S_{fi}$:
\begin{equation} \label{eq:Sfi}
    S_{fi} = \delta_{fi} - i \, (2\pi)^4 \delta^{(4)}(p_f - p_i) M_{fi} \; ,
\end{equation}
where $i$ and $f$ stand for initial and final state, and $p_i$ and $p_f$ for their respective momenta. Eq.~\eqref{eq:Sfi} unambiguously defines the sign of $M_{fi}$, which is necessary in the context of our study. For instance, Eqs.~\eqref{eq:RgiI} and \eqref{eq:RqiI} contain products of amplitudes.

The scattering amplitude, $M_g(\{p_i+\delta_i\},q,\{\sigma_i\},\{c_i\},g^B_s)$, for the process \eqref{eq:process} is given by an expansion in the bare strong coupling constant $g^B_s$:
\begin{equation} \label{eq:Mg}
    M_g \equiv \big( g_s^B )^{n-1} \bigg[ M_g^{(0)} + \frac{\mu^{-2\epsilon}\alpha^B_s}{(4\pi)^{1-\epsilon}} \, M_g^{(1)} + \order{\big( \alpha_s^B \big)^2} \bigg] \; , \qquad \alpha^B_s \equiv \frac{(g^B_s)^2}{4\pi} \; ,
\end{equation}
where $\epsilon$ is the parameter of dimensional regularisation with space-time dimension $d \equiv 4 - 2\epsilon$. Although we work with bare quantities, we have introduced the parameter $\mu$ with unit mass dimension in order to retain the four-dimensional mass dimension of the amplitudes. In what follows, we allow for massive quarks at tree level. Hence, $M^{(0)}_g$ may depend on $m_i \neq 0$. On the other hand, the soft expansion of the one-loop amplitude $M^{(1)}_g$ is only provided in the massless case. The definition of $M_g$ is completed once we assume that the external states are four-dimensional, which corresponds to the `t Hooft-Veltman scheme within the family of dimensional-regularisation schemes.

Finally, the expansion of the {\it reduced scattering amplitude}, $M(\{p_i\},\{\sigma_i\},\{c_i\},g^B_s)$, for the process obtained from \eqref{eq:process} by removing the soft gluon and setting the momentum shifts to zero, is given by:
\begin{equation} \label{eq:M}
\begin{aligned}
    &M \equiv \big( g_s^B )^{n-2} \bigg[ M^{(0)} + \frac{\mu^{-2\epsilon}\alpha^B_s}{(4\pi)^{1-\epsilon}} \, M^{(1)} + \order{\big( \alpha_s^B \big)^2} \bigg] \; .
\end{aligned}
\end{equation}

\subsection{Colour/spin-space formalism}

The soft expansion of Sections~\ref{sec:LBK} and \ref{sec:OneLoopLBK} requires manipulation of the colour state of the hard partons already at $\order{1/\lambda}$. Furthermore, subleading effects at order $\order{\lambda^0}$ require the manipulation of the polarisation state of the hard partons. The formulae are simplified by the use of the colour/spin-space formalism introduced in Ref.~\cite{Catani:1996vz}. This formalism relies on abstract basis vectors:
\begin{equation}
    \ket{c_1,\dots,c_m;\sigma_1,\dots,\sigma_m} \equiv \ket{c_1,\dots,c_m} \otimes \ket{\sigma_1,\dots,\sigma_m} \; ,
\end{equation}
with either $m=n+1$ or $m=n$ in the present case. Accordingly, we define\footnote{The $\mu$ dependence for $l > 0$ is implicit.}:
\begin{equation} \label{eq:MgKet}
    \ket{M_g^{(l)}(\{p_i+\delta_i\},q)} \equiv \sum_{\{\sigma_i\}} \sum_{\{c_i\}} M_g^{(l)}(\{p_i+\delta_i\},q,\{\sigma_i\}\,\{c_i\}) \, \ket{c_1,\dots,c_{n+1};\sigma_1,\dots,\sigma_{n+1}} \; ,
\end{equation}
and similarly for the reduced scattering amplitude:
\begin{equation} \label{eq:Ml}
    \ket{M^{(l)}(\{p_i\})} \equiv \sum_{\{\sigma_i\}} \sum_{\{c_i\}} M^{(l)}(\{p_i\},\{\sigma_i\},\{c_i\}) \, \ket{c_1,\dots,c_n;\sigma_1,\dots,\sigma_n} \; .
\end{equation}
The soft expansion at one-loop order, Eq.~\eqref{eq:OneLoopLBK}, involves {\it flavour off-diagonal} contributions that are identified by a replacement of a pair of partons, $i$ and $j$, in the reduced scattering amplitude w.r.t.\ to the original process \eqref{eq:process}. The replacement does not affect the momenta of the partons. Since we do not introduce a flavour/colour/spin-space in the present publication, the respective reduced amplitude will be denoted by:
\begin{equation}
    \ket{M^{(l)}(\{p_i\}) \, \Big|\substack{a_i \to \tilde{a}_i \\ a_j \to \tilde{a}_j}} \; .
\end{equation}

In order to select amplitudes with a definite polarisation and colour of parton $i$, we define the following surjection operator:
\begin{equation}
    \mathbf{P}_i(\sigma,c) \ket{\dots,c_{i-1},c_i,c_{i+1},\dots;\dots,\sigma_{i-1},\sigma_i,\sigma_{i+1},\dots} \equiv \delta_{\sigma\sigma_i} \delta_{cc_i} \ket{\dots,c_{i-1},c_{i+1},\dots;\dots,\sigma_{i-1},\sigma_{i+1},\dots} \; ,
\end{equation}
and its specialisation:
\begin{equation}
    \mathbf{P}_g(\sigma,c) \equiv \mathbf{P}_{n+1}(\sigma,c) \; .
\end{equation}
Furthermore, we define an operator that exchanges the quantum numbers of $i$ and $j$:
\begin{equation}
    \mathbf{E}_{i,j} \ket{\dots,c_i,\dots,c_j,\dots;\dots,\sigma_i,\dots,\sigma_j,\dots} \equiv \ket{\dots,c_j,\dots,c_i,\dots;\dots,\sigma_j,\dots,\sigma_i,\dots} \; .
\end{equation}

\subsection{Colour operators}

The leading term of the soft expansion is expressed in terms of colour-space operators $\mathbf{T}_i^c$:
\begin{equation}
    \mathbf{T}_i^c \ket{\dots,c_i',\dots} \equiv \sum_{c_i} \\ T^c_{a_i,c_ic_i'} \ket{\dots,c_i,\dots} \; ,
\end{equation}
\begin{equation}
    T^c_{g,ab} = if^{acb} \; , \qquad T^c_{q,ab} = T^c_{ab} \; , \qquad T^c_{\bar{q},ab} = -T^c_{ba} \; .
\end{equation}
The structure constants $f^{abc}$ are defined by $\comm{\mathbf{T}^a_i}{\mathbf{T}^b_j} = i f^{abc} \, \mathbf{T}^c_i \, \delta_{ij}$, while the fun\-da\-men\-tal-rep\-re\-sen\-ta\-tion generators, $T^c_{ab}$, are normalised with $\Tr(T^a T^b) = T_F \delta^{ab}$.

\subsection{Spin operators}

The subleading term of the soft expansion is expressed in terms of spin-space operators $\mathbf{K}^{\mu\nu}_i$:
\begin{equation}
    \mathbf{K}_i^{\mu\nu} \ket{\dots,\sigma_i',\dots} \equiv \sum_{\sigma_i} K^{\mu\nu}_{a_i, \, \sigma_i\sigma_i'}(p_i) \ket{\dots,\sigma_i,\dots} \; ,
\end{equation}
with matrices $K^{\mu\nu}_{a, \, \sigma\sigma'}$ that are anti-symmetric in $\mu,\nu$ and hermitian in $\sigma,\sigma'$:
\begin{equation} \label{eq:Kproperties}
    K^{\mu\nu}_{a, \, \sigma\sigma'} = - K^{\nu\mu}_{a, \, \sigma\sigma'} \; , \qquad K^{\mu\nu \, *}_{a, \, \sigma\sigma'} = K^{\mu\nu}_{a, \, \sigma'\sigma} \; .
\end{equation}
For $p^0 > 0$, i.e.\ for outgoing quarks, anti-quarks and gluons, these matrices are uniquely defined by\footnote{These relations are a consequence of the Lorentz transformation properties of free fields, see for example Section 5.1 of Ref.~\cite{Weinberg:1995mt}.}:
\begin{equation} \label{eq:Kdef}
\begin{aligned}
    &\sum_{\sigma'} K^{\mu\nu}_{q,\sigma\sigma'}(p) \, \bar{u}(p,\sigma') \equiv J^{\mu\nu}(p) \, \bar{u}(p,\sigma) - \frac{1}{2} \bar{u}(p,\sigma) \, \sigma^{\mu\nu} \; , \qquad \sigma^{\mu\nu} \equiv \frac{i}{2} \comm{\gamma^\mu}{\gamma^\nu} \; , \\[.2cm]
    &\sum_{\sigma'} K^{\mu\nu}_{\bar{q},\sigma\sigma'}(p) \, v(p,\sigma') \equiv \Big( J^{\mu\nu}(p) + \frac{1}{2} \sigma^{\mu\nu} \Big) \, v(p,\sigma) \; , \\[.2cm]
    &\sum_{\sigma'} K^{\mu\nu}_{g,\sigma\sigma'}(p) \, \epsilon^*_\alpha(p,\sigma') \equiv \Big( J^{\mu\nu}(p) \, g_{\alpha\beta} + i \big( \delta^\mu_\alpha \delta^\nu_\beta - \delta^\nu_\alpha \delta^\mu_\beta \big) \Big) \, \epsilon^{\beta\,*}(p,\sigma) + \text{terms proportional to } p_\alpha \; ,
\end{aligned}
\end{equation}
where $J^{\mu\nu}(p)$ is the generator of Lorentz transformations for scalar functions of $p$:
\begin{equation}
    J^{\mu\nu}(p) \equiv i \big( p^\mu \partial_p^\nu - p^\nu \partial_p^\mu \big) \; , \qquad \partial_p^\mu \equiv \pdv{p_{\mu}} \; .
\end{equation}
Later, we will mostly use the shorthand notations:
\begin{equation}
    J_i^{\mu\nu} \equiv J^{\mu\nu}(p_i) \; , \qquad \partial_i^\mu \equiv \partial_{p_i}^\mu \; .
\end{equation}

Definitions~\eqref{eq:Kdef} may be rewritten in terms of bi-spinors and polarisation vectors of incoming partons:
\begin{align}
    &\sum_{\sigma'} K^{\mu\nu\,\ast}_{\bar{q},\sigma\sigma'}(p) \, u(p,\sigma') = - \Big( J^{\mu\nu}(p) + \frac{1}{2} \sigma^{\mu\nu} \Big) \, u(p,\sigma) \; , \nonumber \\[.2cm]
    &\sum_{\sigma'} K^{\mu\nu\,\ast}_{q,\sigma\sigma'}(p) \, \bar{v}(p,\sigma') = - \Big( J^{\mu\nu}(p) \, \bar{v}(p,\sigma) - \frac{1}{2} \bar{v}(p,\sigma) \, \sigma^{\mu\nu} \Big) \; , \\[.2cm]
    &\sum_{\sigma'} K^{\mu\nu\,\ast}_{g,\sigma\sigma'}(p) \, \epsilon_\alpha(p,\sigma') =  - \Big( J^{\mu\nu}(p) \, g_{\alpha\beta} + i \big( \delta^\mu_\alpha \delta^\nu_\beta - \delta^\nu_\alpha \delta^\mu_\beta \big) \Big) \, \epsilon^\beta(p,\sigma) + \text{terms proportional to } p_\alpha \; . \nonumber
\end{align}
Due to our process definition~\eqref{eq:process}, negative-energy momenta imply incoming partons. Hence, we define:
\begin{equation}
    K^{\mu\nu}_{a,\sigma\sigma'}(p) \equiv - K^{\mu\nu\,\ast}_{\bar{a},\sigma\sigma'}(-p) = - K^{\mu\nu}_{\bar{a},\sigma'\sigma}(-p) \qquad \text{for} \qquad p^0 < 0 \; .
\end{equation}

Since $v(p,\sigma) = C\bar{u}^T(p,\sigma)$, with $C$ the charge conjugation matrix, the matrices for quarks and anti-quarks fulfil:
\begin{equation}
    K^{\mu\nu}_{\bar{q},\sigma\sigma'}(p) = K^{\mu\nu}_{q,\sigma\sigma'}(p) \; .
\end{equation}
This relation is consistent with the fact that spin and helicity have the same definition for particles and anti-particles.

For a massive-quark bi-spinor, with spin defined in the rest-frame along the third axis, transformed with a pure boost to reach momentum $p$ from $p_0^\mu \equiv (m,\bm{0})$, one finds:
\begin{equation}
    K^{\mu\nu}_{q,\sigma\sigma'} = \frac{\epsilon^{\mu\nu\alpha i} \, \big( p + p_0
  \big)_\alpha}{\big( p + p_0 \big)^0} \, \frac{ \tau^i_{\sigma\sigma'}}{2} \; ,
\end{equation}
where $\tau^i_{\sigma\sigma'}$, $i = 1,2,3$ are the three Pauli matrices.

For massless partons, helicity conservation implies that $K^{\mu\nu}_{a,\sigma\sigma'}$ is proportional to $\delta_{\sigma\sigma'}$. Assuming that bi-spinors and polarisation vectors for the two helicities are related by a momentum-independent anti-linear transformation, one finds:
\begin{equation} \label{eq:Khel}
    K^{\mu\nu}_{a,\sigma\sigma'} = \sigma \, \delta_{\sigma\sigma'} \, K^{\mu\nu} \; .
\end{equation}
Furthermore, it follows from the definitions \eqref{eq:Kdef} that $p_\mu K^{\mu\nu}_{a,\sigma\sigma'} = 0$ for $p^2 = 0$. Hence:
\begin{equation} \label{eq:Keps}
    K^{\mu\nu} = \epsilon^{\mu\nu\alpha\beta} p_\alpha r_\beta \; , \qquad \epsilon_{0123} \equiv +1 \; ,
\end{equation}
for some $r$ that we assume to be lightlike\footnote{If $r^2 \neq 0$, then the replacement $r \to r' \equiv r - r^2 p/2 r\cdot p$ does not change Eq.~\eqref{eq:Keps}, while $r'^2 = 0$.}. In particular, if massless bi-spinors are defined along the third axis and then rotated in the direction of $\bm{p} \equiv E \big(\sin(\theta) \cos(\varphi), \sin(\theta) \sin(\varphi), \cos(\theta) \big) = E R_z(\varphi) R_y(\theta) \bm{\hat{z}}$ with the composition of rotations $R_z(\varphi) R_y(\theta) R_z(-\varphi)$, then:
\begin{equation} \label{eq:MasslessK}
    K^{\mu\nu}(p) = \frac{\epsilon^{\mu\nu\alpha\beta} \, p_\alpha \bar{p}_{0\beta}}{p \cdot \bar{p}_0} \; , \qquad \bar{p}_0^\mu \equiv (E,0,0,-E) \; .
\end{equation}
This result is also valid for polarisation vectors defined in the spinor-helicity formalism using the same bi-spinors:
\begin{equation} \label{eq:EpsSpinHel}
    \epsilon^*_\mu(p,\pm 1) \equiv \pm \frac{\mel{p\pm}{\gamma_\mu}{k\pm}}{\sqrt{2} \ip{k\mp}{p\pm}} \equiv \pm \frac{\bar{u}(p,\pm \tfrac{1}{2}) \, \gamma_\mu \, u(k,\pm \tfrac{1}{2})}{\sqrt{2} \, \bar{u}(k,\mp \tfrac{1}{2}) \, u(p,\pm \tfrac{1}{2})} \; ,
\end{equation}
with an arbitrary lightlike reference vector $k$. If either the massless bi-spinors or the polarisation vectors include an additional phase factor, e.g.\ $\epsilon^{\prime \ast}(p,+1) \equiv \exp(i \phi(p)) \, \epsilon^*(p,+1)$, then $K^{\mu\nu}$ is modified as follows:
\begin{equation} \label{eq:Kprime}
    K^{\prime \, \mu\nu} = K^{\mu\nu} + i J^{\mu\nu} \phi(p) \; .
\end{equation}
With the spinor-helicity-formalism polaristion vectors, direct calculation yields:
\begin{equation} \label{eq:Kcontraction}
    \epsilon_\mu(p,+1) \, \epsilon^*_\nu(p,+1) \, iK^{\mu\nu} = 1\; .
\end{equation}
However, because of \eqref{eq:Kprime}, this result is valid in general. Contractions with $K^{\mu\nu}$ can be efficiently evaluated with the help of:
\begin{equation} \label{eq:Kdecomposition}
    iK_{\mu\nu} = \sum_{\sigma} \text{sgn}(\sigma) \, \epsilon^*_\mu(p,\sigma) \, \epsilon_\nu(p,\sigma) \; ,
\end{equation}
with the polarisation vectors \eqref{eq:EpsSpinHel} assuming $k = r$ and $r$ as in Eq.~\eqref{eq:Keps}.

Besides the spin operator $\mathbf{K}_i^{\mu\nu}$, our results involve a simpler spin-dependent operator that gives the sign of the product of the helicities of parton $i$ and gluon $n+1$:
\begin{equation}
    \mathbf{\Sigma}_{g,i} \ket{\dots,\sigma_i,\dots,\sigma} \equiv \text{sgn}(\sigma\sigma_i)  \ket{\dots,\sigma_i,\dots,\sigma} \; .
\end{equation}

\subsection{Splitting operators} \label{sec:SplittingOperators}

The soft expansion at one-loop order requires the collinear expansion of tree-level amplitudes as stated in Section~\ref{sec:theorem} and proven in Section~\ref{sec:proof}. The latter expansion is expressed in terms of splitting operators that act non-trivially in both colour and spin space. The splitting operators are defined as follows:
\begin{align}
     \mel{c_1,c_2;\sigma_1,\sigma_2}{\mathbf{Split}^{(0)}_{q g \, \leftarrow \, q}(k_1,k_2,k)}{c;\sigma} = -\frac{1}{2 \, k_1 \cdot k_2} \, T^{c_2}_{c_1 c} \, \bar{u}(k_1,\sigma_1) \, \slashed{\epsilon}^*(k_2,\sigma_2) \, u(k,\sigma) \; , \label{eq:SplitQGQ} \\[.2cm]
     \mel{c_1,c_2;\sigma_1,\sigma_2}{\mathbf{Split}^{(0)}_{\bar{q} g \, \leftarrow \, \bar{q}}(k_1,k_2,k)}{c;\sigma} = + \frac{1}{2 \, k_1 \cdot k_2} \, T^{c_2}_{c c_1} \, \bar{v}(k,\sigma) \, \slashed{\epsilon}^*(k_2,\sigma_2) \, v(k_1,\sigma_1) \; , \label{eq:SplitQbarGQbar} \\[.2cm]
     \mel{c_1,c_2;\sigma_1,\sigma_2}{\mathbf{Split}^{(0)}_{q\bar{q} \, \leftarrow \, g}(k_1,k_2,k)}{c;\sigma} = -\frac{1}{2 \, k_1 \cdot k_2} \, T^{c}_{c_1 c_2} \, \bar{u}(k_1,\sigma_1) \, \slashed{\epsilon}(k,\sigma) \, v(k_2,\sigma_2) \; , \label{eq:SplitQQbarG}
\end{align}
\begin{equation} \label{eq:SplitGGG}
\begin{split}
     \qquad \mel{c_1,c_2;\sigma_1,\sigma_2}{\mathbf{Split}^{(0)}_{gg \, \leftarrow \, g}(k_1,k_2,k)}{c;\sigma} = -& \frac{1}{2 \, k_1 \cdot k_2} \, i f^{c_1cc_2} \\[.2cm]
     \times \big( &+ 2k_1 \cdot \epsilon^*(k_2,\sigma_2) \, \epsilon^*(k_1,\sigma_1) \cdot \epsilon(k,\sigma) \\[.2cm]
     \qquad &- 2k_2 \cdot \epsilon^*(k_1,\sigma_1) \, \epsilon^*(k_2,\sigma_2) \cdot \epsilon(k,\sigma) \\[.2cm]
     \qquad &+ (k_2-k_1) \cdot \epsilon(k,\sigma) \, \epsilon^*(k_1,\sigma_1) \cdot \epsilon^*(k_2,\sigma_2) \big) \; .
\end{split}
\end{equation}
In order to simplify the notation, for example in \eqref{eq:DeltaMgDef} and \eqref{eq:HqDef}, we also define the following operator:
\begin{multline} \label{eq:SplitGeneral}
    \mathbf{Split}^{(0)}_{i,n+1 \, \leftarrow \, i}(p_i,p_{n+1},p_i') \ket{\dots,c_i',\dots;\dots,\sigma_i',\dots} = \\[.2cm] \sum_{\sigma_ic_i} \sum_{\sigma_{n+1} c_{n+1}} \mel{c_i,c_{n+1};\sigma_i,\sigma_{n+1}}{\mathbf{Split}^{(0)}_{a_i a_{n+1} \, \leftarrow \, a_i'}(p_i,p_{n+1},p_i')}{c_i';\sigma_i'} \\[.2cm] \times \ket{\dots,c_i,\dots,c_{n+1};\dots,\sigma_i,\dots,\sigma_{n+1}} \; ,
\end{multline}
where $a_i'$ is parton $i$ corresponding to the ket on the left-hand side, and $a_i$, $a_j$ are partons $i$,$j$ corresponding to the ket on the right-hand side of \eqref{eq:SplitGeneral}. In general $a_i' \neq a_i$, as for example in \eqref{eq:HqDef}.

\section{Low-Burnett-Kroll theorem for tree-level QCD} \label{sec:LBK}

The leading and subleading term of the soft expansion, i.e.\ expansion in $\lambda$, of the tree-level amplitude $\ket{M_g^{(0)}(\{p_i+\delta_i\},q,\sigma,c)}$ are given by the QCD generalisation \cite{Casali:2014xpa,Bern:2014vva,Broedel:2014fsa} of the Low-Burnett-Kroll (LBK) theorem
\cite{Low:1958sn,Burnett:1967km} originally proven for QED\footnote{The sign in Eq.~\eqref{eq:LBK2} is a consequence of our convention for the strong coupling constant: we assume that the quark-gluon interaction term in the Lagrangian is $+g^B \bar{q} \slashed{A}^a T^a q$.}:
\begin{align} \label{eq:LBK1}
    &\ket{M_g^{(0)}(\{p_i+\delta_i\},q)} = \mathbf{S}^{(0)}(\{p_i\},\{\delta_i\},q) \, \ket{M^{(0)}(\{p_i\})} + \order{\lambda} \; , \\[.2cm]
    &\mathbf{P}_g(\sigma,c) \, \mathbf{S}^{(0)}(\{p_i\},\{\delta_i\},q) = - \sum_i \mathbf{T}_i^c \otimes \mathbf{S}^{(0)}_i(p_i,\delta_i,q,\sigma) \, \ket{M^{(0)}(\{p_i\})} \; , \label{eq:LBK2} \\[.2cm]
    &\mathbf{S}^{(0)}_i = \frac{p_i \cdot \epsilon^*}{p_i \cdot q} + \frac{1}{p_i \cdot q} \bigg[ \Big( \epsilon^* - \frac{p_i \cdot \epsilon^*}{p_i \cdot q} \, q \Big) \cdot \delta_i + p_i \cdot \epsilon^* \sum_{j} \delta_j \cdot \partial_j + \frac{1}{2} F_{\mu\nu} \Big( J^{\mu\nu}_i - \mathbf{K}^{\mu\nu}_i \Big) \bigg] \; , \label{eq:LBK3}
\end{align}
with:
\begin{equation}
\begin{aligned}
    &\mel{q\sigma c}{A^a_\mu(0)}{0} = \delta^{ca} \epsilon^*(q,\sigma) \; , \\[.2cm]
    &\mel{q\sigma c}{F^a_{\mu\nu}(0)}{0} = \delta^{ca} i \big( q_\mu \epsilon^*_\nu(q,\sigma) - q_\nu \epsilon^*_\mu(q,\sigma) \big) \equiv \delta^{ca} F_{\mu\nu}(q,\sigma) \; ,
\end{aligned}
\end{equation}
where $A^a_\mu(x)$ and $F^a_{\mu\nu}(x)$ are the gluon field and the respective field-strength tensor, while $\ket{q\sigma c}$ is a single-gluon state with momentum $q$, polarisation $\sigma$ and colour $c$.

\begin{figure}
    \centering
    \includegraphics[width=.4\textwidth]{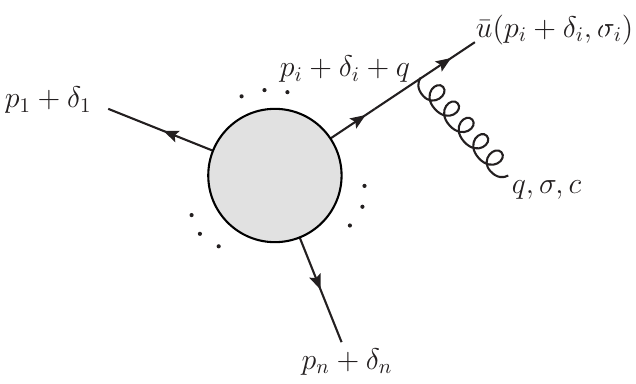}
    \caption{External-emission diagram that yields a contribution to the eikonal approximation in the case of an outgoing quark.}
    \label{fig:eikonal}
\end{figure}

\subsection{Derivation and constraints} \label{sec:LBKderivation}

Most of the terms in Eq.~\eqref{eq:LBK3} are obtained by extending the eikonal approximation to one order higher in $\lambda$. Indeed, consider the diagram of Fig.~\ref{fig:eikonal}. The leading term as well as the first term in the square bracket of Eq.~\eqref{eq:LBK3} are due to the expansion of the eikonal approximation taken with the original momentum, $p_i + \delta_i$, of the hard-parton, i.e.\ outgoing quark in Fig.~\ref{fig:eikonal}:
\begin{equation}
    \frac{(p_i + \delta_i) \cdot \epsilon^*}{(p_i + \delta_i) \cdot q} = \frac{p_i \cdot \epsilon^*}{p_i \cdot q} + \frac{1}{p_i \cdot q} \Big( \epsilon^* - \frac{p_i \cdot \epsilon^*}{p_i \cdot q} \, q \Big) \cdot \delta_i + \order{\lambda} \; .
\end{equation}
The second term in the square bracket in Eq.~\eqref{eq:LBK3} is due to the expansion of the reduced scattering amplitude represented by the shaded circle in Fig.~\ref{fig:eikonal} in $\delta_j$, $j = 1,\dots,n$. The additional expansion of this amplitude in $q$ is taken into account by the first term on the right-hand side of:
\begin{equation} \label{eq:FJ}
    \frac{1}{2 \, p_i \cdot q} F_{\mu\nu} J_i^{\mu\nu} = \frac{p_i \cdot \epsilon^*}{p_i \cdot q} q \cdot \partial_i - \epsilon^* \cdot \partial_i \; .
\end{equation}
The classic LBK argument that generates the second term on the right-hand side of the above equation from the first term on the right-hand side, consists in requiring the soft expansion to fulfil the (QED) Ward identity, i.e.\ transversality of the amplitude with respect to the soft-gluon momentum. This accounts for emissions from the internal off-shell lines, i.e.\ diagrams that do not have the structure of Fig.~\ref{fig:eikonal}.

While spin effects can be obtained by explicit calculation of the expression for Fig.~\ref{fig:eikonal} and similarly for anti-quarks and gluons, there is a simpler argument that allows to understand the result. From Fig.~\ref{fig:eikonal}, we conclude that the external wave function, i.e.\ bi-spinor for quarks and anti-quarks or polarisation vector for gluons, does not depend on $q$. Hence, the differential operator $q \cdot \partial_i$ in Eq.~\eqref{eq:FJ} should not act on it. We thus have to subtract the action of $J^{\mu\nu}$ on the external wave function. The result should, however, still contain a gauge-invariant amplitude with $n$ hard partons. Thus, the subtracted term can be at most a linear combination of amplitudes with different polarisations of the hard parton $a_i$, which leads to the replacement of $J_i^{\mu\nu}$ by $J_i^{\mu\nu} - \mathbf{K}_i^{\mu\nu}$ in Eq.~\eqref{eq:FJ}. The latter difference does not contain any derivatives when acting on external wave functions according to Eqs.~\eqref{eq:Kdef}. This argument has the virtue of applying at higher orders as well. In consequence, the one-loop expression for the soft operator in Eq.~\eqref{eq:Soft} also only contains the combination $J_i^{\mu\nu} - \mathbf{K}_i^{\mu\nu}$.

The soft expansion \eqref{eq:LBK1} is strongly constrained by Lorentz covariance and gauge invariance (Ward identity) as has been discussed in great detail previously in Ref.~\cite{Broedel:2014fsa}, albeit only in the case of pure gluon amplitudes. Here, we would like to stress once more that the process-dependent input on the r.h.s.\ of Eq.~\eqref{eq:LBK1}, i.e.\ the amplitude $\ket{M^{(0)}(\{p_i\})}$, is gauge invariant on its own. This is not a trivial fact, since it does not naively apply in high-energy factorization for example, see Ref.~\cite{Bury:2017jxo} and references therein. In the present case, the issue of gauge invariance is entangled with the issue of defining momentum derivatives in Eq.~\eqref{eq:LBK3}. Indeed, the amplitude $\ket{M^{(0)}(\{p_i\})}$ must be on-shell, and it thus only depends on the spatial components of the momentum vectors. The momentum derivatives in Eq.~\eqref{eq:LBK3}, on the other hand, also involve the energy component. Fortunately, Eq.~\eqref{eq:LBK3}, hence also Eq.~\eqref{eq:LBK1}, is consistent with on-shellness since:
\begin{equation}
    \big( \sum_j \delta_j \cdot \partial_j \big) \, p_i^2 = 2 \, \delta_i \cdot p_i = 0 \; , \qquad J_i^{\mu\nu} \, p_i^2 = 0 \; ,
\end{equation}
where we have used Eq.~\eqref{eq:orthogonality} and neglected terms of higher order in $\lambda$. An additional difficulty arises from the fact that Eq.~\eqref{eq:LBK3} involves derivatives in all of the momenta $p_i$, whereas the amplitude $\ket{M^{(0)}(\{p_i\})}$ is only a function of $n-1$ of them due to momentum conservation. Since extension of $\ket{M^{(0)}(\{p_i\})}$ away from momentum conservation is not unique, Eq.~\eqref{eq:LBK1} must be consistent with momentum conservation. This is indeed the case, albeit colour-conservation is required for the proof:
\begin{equation}
    \bigg[ \mathbf{P}_g(\sigma,c) \mathbf{S}^{(0)}(\{p_i\},\{\delta_i\},q) \bigg]_{\substack{\text{momentum} \\ \text{derivatives}}} \, \ket{f(P)} = \bigg( \epsilon^* \cdot \pdv{P} \bigg) \sum_i \mathbf{T}_i^c \, \ket{f(P)} = 0 \; , \qquad P \equiv \sum_i p_i \; ,
\end{equation}
where $\ket{f(P)}$ is invariant with respect to global gauge transformations and depends on the sum of the momenta only. The importance of this result lies in the fact that the result for the soft expansion in Eq.~\eqref{eq:LBK1} remains the same even if we eliminate one of the $p_i$ momenta in $\ket{M^{(0)}(\{p_i\})}$ by momentum conservation. In fact, one can eliminate different $p_i$'s in different diagrams that contribute to $\ket{M^{(0)}(\{p_i\})}$ without affecting the final result.

\subsection{Squared amplitudes}

While the focus of this publication lies on amplitudes, we would like to point out the simplifications that occur in the case of squared amplitudes summed over spin and colour. The first simplification is the lack of spin effects already noted in Ref.~\cite{Burnett:1967km}. Indeed, squaring Eq.~\eqref{eq:LBK1} and keeping only terms up to $\order{1/\lambda}$, leaves the following contribution containing spin operators:
\begin{equation}
    -i \sum_{ij}\frac{p_i^\mu q^\nu}{p_i \cdot q} \ev{\mathbf{T}_i \cdot \mathbf{T}_j \otimes \big( \mathbf{K}_{i,\mu\nu} - \mathbf{K}^\dagger_{i,\mu\nu} \big)}{M^{(0)}} = 0 \; .
\end{equation}
This contribution vanishes because of the hermiticity, \eqref{eq:Kproperties}, of the spin operators. The second simplification is the possibility \cite{DelDuca:2017twk} to include subleading soft effects through momentum shifts as follows:
\begin{multline} \label{eq:LBKSquared}
    \braket{M^{(0)}_g(\{k_l\},q)} = - \sum_{i \neq j} \left( \frac{k_i \cdot k_j}{(k_i \cdot q)(k_j \cdot q)} - \frac{m_i^2 }{2 \big( k_i \cdot q \big)^2} - \frac{m_j^2}{2 \big( k_j \cdot q \big)^2} \right) \\[.2cm] \times \ev{\mathbf{T}_i \cdot \mathbf{T}_j}{M^{(0)}(\{k_l + \delta_{il} \Delta_i + \delta_{jl} \Delta_j\})} + \order{\lambda^0} \; ,
\end{multline}
with:
\begin{equation}
\begin{aligned}
    k_i &\equiv p_i + \delta_i \; , \\[.2cm]
    \Delta_i &\equiv \frac{1}{N_{ij}} \left[ \left( 1 - \frac{m_i^2 \big( p_j \cdot q \big)}{\big( p_j \cdot p_i \big) \big( p_i \cdot q \big)} \right) q + \frac{p_j \cdot q}{p_j \cdot p_i} \, p_i - \frac{p_i \cdot q}{p_i \cdot p_j} \, p_j \right] \; , \\[.2cm]
    \Delta_j &\equiv \frac{1}{N_{ij}} \left[ \left( 1 - \frac{m_j^2 \big( p_i \cdot q \big)}{\big( p_i \cdot p_j \big) \big( p_j \cdot q \big)} \right) q - \frac{p_j \cdot q}{p_i \cdot p_j} \, p_i + \frac{p_i \cdot q}{p_i \cdot p_j} \, p_j \right] \; , \\[.2cm]
    N_{ij} &\equiv 2 - \frac{m_i^2 \big( p_j \cdot q \big)}{\big( p_j \cdot p_i \big) \big( p_i \cdot q \big)} - \frac{m_j^2 \big( p_i \cdot q \big)}{\big( p_i \cdot p_j \big) \big( p_j \cdot q \big)} \; .
\end{aligned}
\end{equation}
Notice that the momenta in the reduced scattering amplitude in Eq.~\eqref{eq:LBKSquared} satisfy momentum conservation and are on-shell up to  $\order{\lambda}$:
\begin{equation}
    \sum_{l} k_l + \delta_{il} \Delta_i + \delta_{jl} \Delta_j = 0 \; , \qquad \big( k_l + \delta_{il} \Delta_i + \delta_{jl} \Delta_j \big)^2 = m_l^2 + \order{\lambda^2} \; .
\end{equation}
In fact, it is possible to add corrections of $\order{\lambda^2}$ to these momenta to make them exactly on-shell.

\section{Soft expansion of massless one-loop QCD amplitudes} \label{sec:OneLoopLBK}

\subsection{Theorem} \label{sec:theorem}

The main result of this publication is the following next-to-leading-power-accurate soft expansion of a one-loop massless-QCD amplitude:
\begin{equation} \label{eq:OneLoopLBK}
\begin{split}
    &\Big| M_g^{(1)}(\{p_i+\delta_i\},q) \Big\rangle = \mathbf{S}^{(0)}(\{p_i\},\{\delta_i\},q) \, \ket{M^{(1)}(\{p_i\})} \\[.2cm]
    &\quad+ \mathbf{S}^{(1)}(\{p_i\},\{\delta_i\},q) \, \ket{M^{(0)}(\{p_i\})} + \int_0^1 \dd{x} \sum_i \mathbf{J}_i^{(1)}(x,p_i,q) \ket{H^{(0)}_{g,i}(x,\{p_i\},q)} \\[.2cm]
    &\quad+ \sum_{i \neq j} \sum_{\substack{\tilde{a}_i \neq a_i \\ \tilde{a}_j \neq a_j}} \mathbf{\tilde{S}}^{(1)}_{a_i a_j \, \leftarrow \, \tilde{a}_i \tilde{a}_j, \, ij}(p_i,p_j,q) \, \ket{M^{(0)}(\{p_i\}) \, \Big|{\substack{a_i \, \to \, \tilde{a}_i \\ a_j \, \to  \, \tilde{a}_j}}} + \int_0^1 \dd{x} \sum_{\substack{i \\ a_i = g}} \mathbf{\tilde{J}}_i^{(1)}(x,p_i,q) \ket{H^{(0)}_{\bar{q},i}(x,\{p_i\},q)} \\[.2cm]
    &\quad+ \order{\lambda} \; .
\end{split}
\end{equation}

The {\it soft operator} $\mathbf{S}^{(1)}(\{p_i\},\{\delta_i\},q)$ is an extension of the one-loop soft current, and is given by the expansion through $\order{\lambda^0}$ of the r.h.s.\ of:
\begin{multline} \label{eq:Soft}
    \mathbf{P}_g(\sigma,c) \, \mathbf{S}^{(1)}(\{p_i\},\{\delta_i\},q) + \order{\lambda} = \frac{2 \, r_{\text{Soft}}}{\epsilon^2} \, \sum_{i \neq j} i f^{abc} \mathbf{T}_i^a \mathbf{T}_j^b \otimes \Bigg(- \frac{\mu^2 s^{(\delta)}_{ij}}{s^{(\delta)}_{iq} s^{(\delta)}_{jq}} \Bigg)^\epsilon \Bigg[ \textbf{S}^{(0)}_i(p_i,\delta_i,q,\sigma) \\[.2cm]
    + \frac{\epsilon}{1-2\epsilon} \frac{1}{p_i \cdot p_j} \Bigg( \frac{p_i^\mu p_j^\nu - p_j^\mu p_i^\nu}{p_i \cdot q} + \frac{p_j^\mu p_j^\nu}{p_j \cdot q} \Bigg) F_{\mu\rho}(q,\sigma) \, \big( J_i - \mathbf{K}_i \big)_\nu^{\phantom{\nu}\rho} \Bigg] \; ,
\end{multline}
with:
\begin{equation} \label{eq:invariantsDelta}
    s^{(\delta)}_{ij} \equiv 2 \, (p_i + \delta_i) \cdot (p_j + \delta_j) + i0^+ \; , \qquad s^{(\delta)}_{iq} \equiv 2 \, (p_i + \delta_i) \cdot q + i0^+ \; , \qquad s^{(\delta)}_{jq} \equiv 2 \, (p_j + \delta_j) \cdot q + i0^+ \; ,
\end{equation}
\begin{equation} \label{eq:rSoft}
    r_{\text{Soft}} \equiv \frac{\Gamma^3(1 - \epsilon) \Gamma^2(1 + \epsilon)}{\Gamma(1 - 2 \epsilon)} = 1 + \order{\epsilon} \; .
\end{equation}
For convenience, we have not expanded the factor containing $s^{(\delta)}_{ij}$, $s^{(\delta)}_{iq}$ and $s^{(\delta)}_{jq}$. A strict expansion depends on:
\begin{equation} \label{eq:invariants}
    s_{ij} \equiv 2 \, p_i  \cdot p_j + i0^+ \; , \qquad s_{iq} \equiv 2 \, p_i \cdot q + i0^+ \; , \qquad s_{jq} \equiv 2 \, p_j \cdot q + i0^+ \; ,
\end{equation}
and on the scalar products of $\delta_i$ and $\delta_j$ with $p_i$, $p_j$ and $q$. Finally, we notice that contractions of $\mathbf{K}_i^{\mu\nu}$ with other vectors can be conveniently evaluated with the help of Eq.~\eqref{eq:Kdecomposition}.

\begin{figure}
    \centering
    \includegraphics[width=\textwidth]{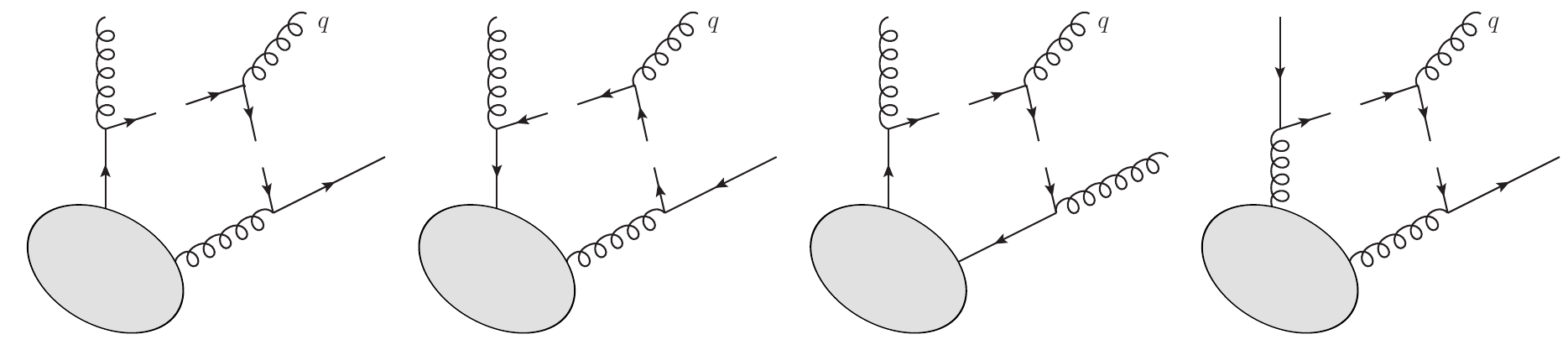}
    \caption{Flavour-off-diagonal contributions described by the operator \eqref{eq:SoftTilde}.}
    \label{fig:flavour-off-diagonal-soft}
\end{figure}

The {\it flavour-off-diagonal soft operator} is given by:
\begin{equation} \label{eq:SoftTilde}
\begin{split}
    &\mathbf{\tilde{S}}^{(1)}_{a_i a_j \, \leftarrow \, \tilde{a}_i \tilde{a}_j, \, ij}(p_i,p_j,q) \, \ket{\dots,c_i',\dots,c_j',\dots;\dots,\sigma_i',\dots,\sigma_j',\dots} \\[.2cm]
    &= - \frac{r_{\text{Soft}}}{\epsilon(1-2\epsilon)} \Bigg(- \frac{\mu^2 s_{ij}}{s_{iq} s_{jq}} \Bigg)^\epsilon \sum_{\sigma c} \sum_{\sigma_ic_i} \sum_{\sigma_jc_j} \sum_{\sigma_i''c_i''} \sum_{\sigma_j''c_j''} \\[.2cm]
    &\quad \begin{cases} T^c_{c_j''c_i''} \, \bar{v}(p_j,\sigma_j'') \, \slashed{\epsilon}^*(q,p_i,\sigma) \, u(p_i,\sigma_i'') & \text{for $a_i = q$ or $\tilde{a}_i = \bar{q}$} \\[.2cm] T^c_{c_i''c_j''} \, \bar{v}(p_i,\sigma_i'') \, \slashed{\epsilon}^*(q,p_i,\sigma) \, u(p_j,\sigma_j'') & \text{for $a_i = \bar{q}$ or $\tilde{a}_i = q$} \end{cases} \\[.2cm]
    &\quad \times \mel{c_i,c_j'';\sigma_i,\sigma_j''}{\mathbf{Split}^{(0)}_{a_i \tilde{\tilde{a}}_j \, \leftarrow \, \tilde{a}_i}(p_i,p_j,p_i)}{c_i';\sigma_i'} \mel{c_j,c_i'';\sigma_j,\sigma_i''}{\mathbf{Split}^{(0)}_{a_j \tilde{\tilde{a}}_i \, \leftarrow \, \tilde{a}_j}(p_j,p_i,p_j)}{c_j';\sigma_j'} \\[.2cm]
    &\quad \times \ket{\dots,c_i,\dots,c_j,\dots,c;\dots,\sigma_i,\dots,\sigma_j,\dots,\sigma} \; ,
\end{split}
\end{equation}
where:
\begin{equation}
    \epsilon^*_\mu(q,p_i,\sigma) \equiv \epsilon^*_\mu(q,\sigma) - \frac{p_i \cdot \epsilon^*(q,\sigma)}{p_i \cdot q} \, q_\mu = i F_{\mu\nu}(q,\sigma) \frac{p_i^\nu}{p_i \cdot q} \; , \qquad \epsilon^*(q,p_i,\sigma) \cdot q = \epsilon^*(q,p_i,\sigma) \cdot p_i = 0 \; .
\end{equation}
The partons $\tilde{\tilde{a}}_i$ and $\tilde{\tilde{a}}_j$ are uniquely determined by flavour conservation in the splitting processes $a_i \tilde{\tilde{a}}_j \, \leftarrow \, \tilde{a}_i$ and $a_j \tilde{\tilde{a}}_i \, \leftarrow \, \tilde{a}_j$. The contribution corresponds to the emission of a soft quark-anti-quark pair, which then produces the soft gluon as depicted in Fig.~\ref{fig:flavour-off-diagonal-soft}. Finally, due to chirality and angular-momentum conservation, we notice:
\begin{equation}
    \text{sgn}(\sigma_i) = \text{sgn}(\sigma_i') = \text{sgn}(\sigma_i'') = - \text{sgn}(\sigma_j'') = -\text{sgn}(\sigma_j') = - \text{sgn}(\sigma_j) \; .
\end{equation}

The {\it jet operator} $\mathbf{J}_i^{(1)}(x,p_i,q)$ is given by:
%
% Keep original with explicit indices
%
%\begin{equation}
%\begin{split}
%    \mathbf{J}_i^{(1)}(x,p_i,q) & \ket{\dots,c_i',\dots,c';\dots,\sigma_i',\dots,\sigma'} = \\[.2cm]
%    &\frac{\Gamma(1+\epsilon)}{1-\epsilon} \bigg( - \frac{\mu^2}{s_{iq}} \bigg)^{\epsilon} \big( x(1-x) \big)^{-\epsilon} \sum_{cc_i} \bigg( T^c_{a_i} T^{c'}_{a_i} + \frac{1}{x} i f^{cdc'} T^d_{a_i} \bigg)_{c_ic_i'} \\[.2cm] &\times \sum_{\sigma\sigma_i} \epsilon^*_\mu(q,p_i,\sigma) \epsilon_\nu(p_i,\sigma') \big( (x - 2) \, g^{\mu\nu} \, \delta_{\sigma_i\sigma_i'} + \big( 1 + 2\dim(a_i) \big) \, x \, iK_{a_i,\sigma_i\sigma_i'}^{\mu\nu}(p_i) \big) \\[.2cm]
%    &\times \ket{\dots,c_i,\dots,c;\dots,\sigma_i,\dots,\sigma} \; ,
%\end{split}
%\end{equation}
%
\begin{equation} \label{eq:Jet}
\begin{split}
    \mathbf{P}_g(\sigma,c) \, \mathbf{J}_i^{(1)}(x,p_i,q) &= \frac{\Gamma(1+\epsilon)}{1-\epsilon} \bigg( - \frac{\mu^2}{s_{iq}} \bigg)^{\epsilon} \big( x(1-x) \big)^{-\epsilon} \sum_{\sigma'c'} \epsilon^*_\mu(q,p_i,\sigma) \epsilon_\nu(p_i,\sigma') \, \mathbf{P}_g(\sigma',c') \\[.2cm]
    &\qquad \times \bigg[ \bigg( \mathbf{T}^{c}_i \mathbf{T}^{c'}_i + \frac{1}{x} i f^{cdc'} \mathbf{T}^d_i \bigg) \otimes \big( (x - 2) \, g^{\mu\nu} + \big( 1 + 2\dim(a_i) \big) \, x \, i \mathbf{K}_i^{\mu\nu} \big) \bigg] \\[.2cm]
    &= \frac{\Gamma(1+\epsilon)}{1-\epsilon} \bigg( - \frac{\mu^2}{s_{iq}} \bigg)^{\epsilon} \big( x(1-x) \big)^{-\epsilon} \epsilon^*(q,p_i,\sigma) \cdot \epsilon(p_i,-\sigma) \sum_{c'} \mathbf{P}_g(-\sigma,c') \\[.2cm]
    &\qquad \times \bigg[ \bigg( \mathbf{T}^{c}_i \mathbf{T}^{c'}_i + \frac{1}{x} i f^{cdc'} \mathbf{T}^d_i \bigg) \otimes \big( - 2 + x \big( 1 + \mathbf{\Sigma}_{g,i} \big) \big) \bigg] \; ,
\end{split}
\end{equation}
where $\dim(a_i)$ is the mass dimension of the wave function of parton $i$, $\dim(q) = \dim(\bar{q}) = \sfrac{1}{2}$ and $\dim(g) = 0$. The second equality follows from Eqs.~\eqref{eq:Khel} and \eqref{eq:Kcontraction}:
\begin{equation} \label{eq:KforJet}
    \epsilon^*_\mu(q,p_i,\sigma) \epsilon_\nu(p_i,\sigma') \, iK^{\mu\nu}_{a_i,\sigma_i\sigma_i'}(p_i) = - \sigma \delta_{-\sigma\sigma'} \, \sigma_i \delta_{\sigma_i\sigma_i'} \, \epsilon^*(q,p_i,\sigma) \cdot \epsilon(p_i,-\sigma) \; ,
\end{equation}
because $\epsilon^*(q,p_i,\sigma)$ has helicity $\sigma$ as a polarisation vector for $q$ and helicity $-\sigma$ as a polarisation vector for $p_i$. This can be proven in the rest-frame of $q+p_i$, where a clockwise rotation around $q$ is equivalent to an anti-clockwise rotation around $p_i$.

The jet operator of a gluon, $a_i = g$, is not symmetric w.r.t.\ the gluons $i$ and $n+1$. On the other hand it is given by the same expression as that of the (anti-)quark up to the factor depending on $\dim(a_i)$. In fact, because of Eq.~\eqref{eq:Khel}, the spin-dependent parts of the (anti-)quark and gluon jet operators are numerically identical. This is not a coincidence, but rather a consequence of a hidden supersymmetry. Indeed, if the quark field transformed with the adjoint representation of the gauge group, then it could belong to the same superfield as the gluon, and the diagrams that enter the calculation of the jet operator for a quark and for a gluon would be related by supersymmetry. The missing symmetry of the gluon jet operator, on the other hand, is restored in the convolution with the symmetric collinear-gluon amplitude \eqref{eq:HgDefGluon}.

The {\it flavour-off-diagonal jet operator} $\mathbf{\tilde{J}}_i^{(1)}(x,p_i,q)$ is given by:
\begin{equation} \label{eq:JetTilde}
\begin{split}
    &\mathbf{\tilde{J}}_i^{(1)}(x,p_i,q) \, \ket{\dots,c_i',\dots,c';\dots,\sigma_i',\dots,\sigma'} = \\[.2cm]
    &\quad\;\; \frac{\Gamma(1+\epsilon)}{1-\epsilon} \bigg( - \frac{\mu^2}{s_{iq}} \bigg)^{\epsilon} \big( x(1-x) \big)^{-\epsilon} \sum_{cc_i} \bigg( T^c_q T^{c_i}_q + x i f^{cdc_i} T^d_q \bigg)_{c'c_i'} \sum_{\sigma\sigma_i} \epsilon^*_\mu(q,p_i,\sigma) \epsilon^*_\nu(p_i,\sigma_i) \\[.2cm]
    &\qquad \times \big( ( 1 - 2 x ) \, g^{\mu\nu} \, \mathbbm{1} + 2 \, iK_q^{\mu\nu}(p_i) \big)_{-\sigma'\sigma_i'} \ket{\dots,c_i,\dots,c;\dots,\sigma_i,\dots,\sigma} \\[.2cm]
    &= \frac{\Gamma(1+\epsilon)}{1-\epsilon} \bigg( - \frac{\mu^2}{s_{iq}} \bigg)^{\epsilon} \big( x(1-x) \big)^{-\epsilon} \sum_{cc_i} \bigg( T^c_q T^{c_i}_q + x i f^{cdc_i} T^d_q \bigg)_{c'c_i'} \delta_{-\sigma'\sigma_i'} \sum_{\sigma\sigma_i} \delta_{\sigma\sigma_i} \epsilon^*(q,p_i,\sigma) \cdot \epsilon^*(p_i,\sigma_i) \\[.2cm]
    &\qquad \times \big( - 2 x  + 1 + \text{sgn}(\sigma_i\sigma') \big) \ket{\dots,c_i,\dots,c;\dots,\sigma_i,\dots,\sigma}\; .
\end{split}
\end{equation}
The operator transforms a state with $a_i = q$, $a_{n+1} = \bar{q}$ into a state with $a_i = a_{n+1} = g$. The sign of the r.h.s.\ of Eq.~\eqref{eq:JetTilde} is a consequence of our convention:
\begin{equation} \label{eq:UVrelation}
    v(p,\sigma) = -u(p,-\sigma) \; ,
\end{equation}
see Eq.~\eqref{eq:JtildeSign}. We point out that there is a crossing-like relation between $\mathbf{J}_i$ and $\mathbf{\tilde{J}}_i$ which becomes apparent by comparing the r.h.s.\ of \eqref{eq:JetTilde} with $x \mathbf{J}_i(1/x,p_i,q)$ at vanishing $\epsilon$.

The {\it collinear-gluon amplitude} $\ket*{H^{(0)}_{g,i}(x,\{p_i\},q)}$ is defined as follows for $a_i \in \{q,\bar{q}\}$:
\begin{multline}  \label{eq:HgDefQuark}
    \mathbf{P}_g(\sigma,c) \ket{H^{(0)}_{g,i}(x,\{p_i\},q)} \equiv \\[.2cm]
    (1-x)^{-\dim(a_i)} \mathbf{P}_g(\sigma,c) \ket{\Delta M^{(0)}_g(x,\{p_i\},q)} - \frac{1}{x} \frac{q \cdot \epsilon^*(p_i,\sigma)}{q \cdot p_i} \mathbf{T}_i^c \ket{M^{(0)}(\{p_i\})}\; ,
\end{multline}
and as follows for $a_i = g$:
\begin{equation} \label{eq:HgDefGluon}
\begin{split}
    \mathbf{P}_i(\sigma_i,c_i) \mathbf{P}_{n+1}(\sigma_{n+1},c_{n+1}) &\ket{H^{(0)}_{g,i}(x,\{p_i\},q)} \equiv \\[.2cm]
    &(1-x)^{-\dim(a_i)} \mathbf{P}_i(\sigma_i,c_i) \mathbf{P}_{n+1}(\sigma_{n+1},c_{n+1}) \ket{\Delta M^{(0)}_g(x,\{p_i\},q)} \\[.2cm]
    &- \frac{1}{x} \frac{q \cdot \epsilon^*(p_i,\sigma_{n+1})}{q \cdot p_i} \mathbf{P}_i(\sigma_i,c_i) \mathbf{T}_i^{c_{n+1}} \ket{M^{(0)}(\{p_i\})} \\[.2cm]
    &- \frac{1}{1-x} \frac{q \cdot \epsilon^*(p_i,\sigma_i)}{q \cdot p_i} \mathbf{P}_i(\sigma_{n+1},c_{n+1}) \mathbf{T}_i^{c_i} \ket{M^{(0)}(\{p_i\})} \; ,
\end{split}
\end{equation}
where:
\begin{equation} \label{eq:DeltaMgDef}
    \ket{\Delta M^{(0)}_{g,i}(x,\{p_i\},q)} \equiv \lim_{l_\perp \to 0} \bigg[ \ket{M_g^{(0)}(\{k_i\}_{i=1}^n,k_g)} - \mathbf{Split}^{(0)}_{i,n+1 \, \leftarrow \, i}(k_i,k_g,p_i) \ket{M^{(0)}(\{p_i\})} \bigg] \; ,
\end{equation}
is the subleading term of the expansion of the tree-level soft-gluon emission amplitude in the limit of the soft gluon collinear to parton $i$ as specified by the following configuration:
\begin{align}
    &k_g \equiv x p_i + l_\perp - \frac{l_\perp^2}{2x} \frac{q}{p_i \cdot q} \; , & \text{with} & \qquad l_\perp \cdot p_i = l_\perp \cdot q = 0 \; , \\[.2cm]
    & k_i \equiv (1-x) p_i - l_\perp - \frac{l_\perp^2}{2(1-x)} \frac{q}{p_i \cdot q} \; , & \text{and}
    & \qquad k_j \equiv p_j + \order{l_\perp^2} \; , \qquad j \neq i \; .
\end{align}
For $a_i = g$, we further require that the gluon polarisation vector in the amplitude for the subtraction term and hence also in the splitting operator in \eqref{eq:DeltaMgDef} be defined with {\it reference vector} $q$ yielding the helicity sum:
\begin{equation} \label{eq:HelSumSplit}
    \sum_\sigma \epsilon_\mu(p_i,\sigma) \epsilon^*_\nu(p_i,\sigma) = - g_{\mu\nu} + \frac{p_{i\mu} q_\nu + p_{i\nu} q_\mu}{p_i \cdot q} \; .
\end{equation}
Without this requirement, the collinear-gluon amplitude depends on the additional reference vector. Notice that the subtraction in \eqref{eq:DeltaMgDef} removes not only the leading collinear-singular asymptotics, but also part of the regular $\order{l_\perp^0}$ term. The additional term in Eq.~\eqref{eq:HgDefGluon} w.r.t.\ \eqref{eq:HgDefQuark} is necessary in order to retain symmetry w.r.t.\ to the exchange of the gluons $i$ and $n+1$.

The {\it collinear-quark amplitude} $\ket*{H^{(0)}_{\bar{q},i}(x,\{p_i\},q)}$ is given by:
\begin{multline} \label{eq:HqDef}
      \ket{H^{(0)}_{\bar{q},i}(x,\{p_i\},q)} \equiv \\[.2cm] \big( x (1-x) \big)^{-\sfrac{1}{2}} \lim_{l_\perp \to 0} \bigg[ \ket{M_{\bar{q}}^{(0)}(\{k_i\}_{i=1}^n,k_g) \, \Big|_{a_i \to q}} - \mathbf{Split}^{(0)}_{i,n+1 \, \leftarrow \, i}(k_i,k_g,p_i) \ket{M^{(0)}(\{p_i\})} \bigg] \; ,
\end{multline}
where $\ip*{c_1,\dots,c;\sigma_1,\dots,\sigma}{M_{\bar{q}}^{(0)}(\{k_i\}_{i=1}^n,k_g) \, |_{a_i \to \bar{q}}}$ is the amplitude for the process:
\begin{equation}
    0 \to a_1(k_1, \sigma_1, c_1) + \dots + q(k_i, \sigma_i, c_i) + \dots + a_n(k_n, \sigma_n, c_n) + \bar{q}(k_g, \sigma_{n+1}, c_{n+1}) \; .
\end{equation}
If there is more than one massless quark flavour, then the last term in Eq.~\eqref{eq:OneLoopLBK} includes summation over flavours.

\noindent
The {\it collinear convolutions}, i.e.\ integrals over $x$, in Eq.~\eqref{eq:OneLoopLBK} are evaluated explicitly in Section~\ref{sec:convolutions}.

\subsection{Collinear amplitudes} \label{sec:CollinearAmplitudes}

Although Eq.~\eqref{eq:OneLoopLBK} involves convolutions of jet operators with collinear amplitudes, the $x$-integrals can be performed analytically which yields an expression in terms of tree-level amplitudes independent of $x$. In order to derive the relevant formulae, we first list the properties of the collinear amplitudes.

\subsubsection*{Gauge invariance and Ward identity}

By construction, $\ket{\Delta M^{(0)}_{g,i}(x,\{p_i\},q)}$ defined in Eq.~\eqref{eq:DeltaMgDef} is gauge invariant, since it only involves gauge invariant amplitudes. However, it does not satisfy the naive Ward identity w.r.t.\ to the gluon with momentum $x p_i$. If we denote by $s$ the scalar polarisation, i.e.\ $\epsilon^*(p,\sigma = s) = p$, then:
\begin{multline}
    \lim_{l_\perp \to 0} \mathbf{P}_g(\sigma = s, c) \bigg[ \ket{M_g^{(0)}(\{k_i\}_{i=1}^n,k_g)} - \mathbf{Split}^{(0)}_{i,n+1 \, \leftarrow \, i}(k_i,k_g,p_i) \ket{M^{(0)}(\{p_i\})} \bigg] = \\[.2cm] (1-x)^{\dim(a_i)} \mathbf{T}_i^c \ket{M^{(0)}(\{p_i\})} \; .
\end{multline}
The result is entirely due to the second term in the square bracket. It follows that the collinear-gluon amplitudes defined in Eqs.~\eqref{eq:HgDefQuark} and \eqref{eq:HgDefGluon} satisfy the Ward identity:
\begin{equation}
    \mathbf{P}_g(\sigma = s, c) \ket{H^{(0)}_{g,i}(x,\{p_i\},q)} = 0 \; .
\end{equation}

\subsubsection*{Evaluation for arbitrary $x$}

The limit in the definition \eqref{eq:DeltaMgDef} can be obtained directly from Feynman diagrams as follows:
\begin{equation} \label{eq:DeltaMgDirect}
\begin{split}
    \mathbf{P}_i(\sigma_i,c_i) & \mathbf{P}_g(\sigma,c) \ket{\Delta M^{(0)}_{g,i}(x,\{p_i\},q)} = \\[.2cm]
    &\bigg[ \mathbf{P}_i(\sigma_i,c_i) \mathbf{P}_g(\sigma,c) \ket{M_g^{(0)}(\{p_1,\dots,(1-x)p_i,\dots,p_n\},xp_i)} \bigg]_{\substack{\text{non-singular} \\ \text{diagrams}}} \\[.2cm]
    &- \delta_{\sigma_i,-s_i \sigma} \sum_{c_i'} T^c_{a_i,c_ic_i'} \left[ 
    \begin{dcases} \frac{\bar{u}\big((1-x)p_i,\sigma_i\big) \slashed{\epsilon}^*(p_i,\sigma)\slashed{q}}{2\, p_i \cdot q} \pdv{\bar{u}_i} & \text{if $a_i = q$} \\[.4cm]
    \frac{\slashed{q}\slashed{\epsilon}^*(p_i,\sigma)v\big((1-x)p_i,\sigma_i\big)}{2\, p_i \cdot q}  \pdv{v_i} & \text{if $a_i = \bar{q}$} \\[.4cm]
    \frac{(2x-1) q}{p_i \cdot q} \cdot \pdv{\epsilon^*_{i}} & \text{if $a_i = g$}
    \end{dcases}
    \right]
    \mathbf{P}_i(\sigma_i,c_i') \ket{M^{(0)}(\{p_i\})} \; ,
\end{split}
\end{equation}
where $s_i = \sfrac{1}{2}$ if either $a_i = q$ or $a_i = \bar{q}$, and $s_i = 1$ if $a_i = g$. The derivatives $\pdv*{\psi_i}$, $\psi_i \in \{\bar{u}_i,v_i,\epsilon^*_i\}$, remove the wave function $\psi_i$ of parton $i$ in the amplitude. The collinear-quark amplitude is obtained similarly:
\begin{equation} \label{eq:HqDirect}
\begin{split}
    \mathbf{P}_i(\sigma_i,c_i) & \mathbf{P}_g(\sigma,c) \ket{H^{(0)}_{\bar{q},i}(x,\{p_i\},q)} = \\[.2cm]
    &\big( x (1-x) \big)^{-\sfrac{1}{2}} \, \bigg[ \mathbf{P}_i(\sigma_i,c_i) \mathbf{P}_g(\sigma,c) \ket{M_{\bar{q}}^{(0)}(\{p_1,\dots,(1-x)p_i,\dots,p_n\},xp_i)} \, \Big|_{a_i \to q} \bigg]_{\substack{\text{non-singular} \\ \text{diagrams}}} \\[.2cm]
    &- \delta_{\sigma_i,-\sigma} \sum_{c_i'} T^{c_i'}_{c_ic} \, \frac{2q}{p_i \cdot q} \cdot \pdv{\epsilon^*_{i}} \, \mathbf{P}_i(\sigma_i,c_i') \ket{M^{(0)}(\{p_i\})} \; .
\end{split}
\end{equation}

\subsubsection*{Small-$x$ expansion}

\begin{multline} \label{eq:HgSmallX}
    \mathbf{P}_g(\sigma,c) \ket{H^{(0)}_{g,i}(x,\{p_i\},q)} = - \sum_{j \neq i} \mathbf{T}_j^c \otimes \bigg[ \bigg( \frac{1}{x} + \dim(a_i) \bigg) \bigg( \frac{p_j \cdot \epsilon_i^*}{p_j \cdot p_i} - \frac{q \cdot \epsilon_i^*}{q \cdot p_i} \bigg) \\[.2cm]
    + \frac{F_{i\,\mu\nu}}{2 \, p_j \cdot p_i} \Big( - i \big( p_j^\mu \partial_i^\nu - p_j^\nu \partial_i^\mu \big) + J_j^{\mu\nu} - \mathbf{K}_j^{\mu\nu} \Big) + \frac{i q_\mu \epsilon_{i\,\nu}^*}{q \cdot p_i} \, \mathbf{K}_i^{\mu\nu} \bigg] \ket{M^{(0)}(\{ p_i \})} + \order{x} \; ,
\end{multline}
where:
\begin{equation}
    \epsilon_i^* \equiv \epsilon^*(p_i,\sigma) \; , \qquad F_i^{\mu\nu} \equiv i \big( p_i^\mu \epsilon_i^{\nu \, *} - p_i^\nu \epsilon_i^{\mu \, *} \big) \; .
\end{equation}
The above result can be obtained similarly to Eq.~\eqref{eq:LBK3} by extending the eikonal approximation of Eq.~\eqref{eq:DeltaMgDirect} for soft-gluon emission from partons $j \neq i$ with $\delta_k = -\delta_{ki} \, x p_i$ and $q = x p_i$. Subsequently requiring the Ward identity to be satisfied introduces the term:
\begin{equation}
    - \sum_{j \neq i} \mathbf{T}_j^c \, \epsilon^*_i \cdot (\partial_i - \partial_j) \; .
\end{equation}
Spin effects for partons $j \neq i$ are restored as discussed in Section~\ref{sec:LBK}. Finally, contributions due to soft-gluon emission from parton $i$ are given explicitly in Eqs.~\eqref{eq:HgDefQuark} and \eqref{eq:HgDefGluon}, while spin effects can be determined from Eq.~\eqref{eq:DeltaMgDirect}.

\subsubsection*{Dependence on $x$}

It follows from the definitions Eqs.~\eqref{eq:HgDefQuark}, \eqref{eq:HgDefGluon} together with Eq.~\eqref{eq:DeltaMgDirect} evaluated in Feynman gauge that the collinear-gluon amplitudes are not only rational in $x$ but can be reduced by partial fractioning to the form:
\begin{equation} \label{eq:HgDecomposition}
\begin{split}
    \ket{H^{(0)}_{g,i}(x,\{p_i\},q)} &= \bigg( \frac{1}{x} + \dim(a_i) \bigg) \ket{S^{(0)}_{g,i}(\{p_i\},q)} + \ket{C^{(0)}_{g,i}(\{p_i\},q)} + \frac{x}{1-x} \ket{\bar{S}^{(0)}_{g,i}(\{p_i\},q)} \\[.2cm]
    &\quad + \sum_I \bigg( \frac{1}{x_I - x} - \frac{1}{x_I} \bigg) \ket{R^{(0)}_{g,i,I}(\{p_i\})} + x \ket{L^{(0)}_{g,i}(\{p_i\},q)} \; ,
\end{split}
\end{equation}
where the sum in the second line is taken over subsets:
\begin{equation}
    I \subset \{ 1,\dots,n \} \setminus \{i\} \; , \qquad 2 \leq |I| < n-2 \; ,
\end{equation}
with:
\begin{equation}
    \qquad x_I \equiv - \frac{P_I^2 + i0^+}{2 \, p_i \cdot P_I} \; , \qquad P_I \equiv \sum_{j \in I} p_j \; .
\end{equation}

The {\it soft-pole} and {\it constant} contributions, $\ket*{S^{(0)}_{g,i}(\{p_i\},q)}$ and $\ket*{C^{(0)}_{g,i}(\{p_i\},q)}$, follow from Eq.~\eqref{eq:HgSmallX}:
\begin{align} \label{eq:Sgi}
    \mathbf{P}_g(\sigma,c)& \, \ket{S^{(0)}_{g,i}(\{p_i\},q)} = - \sum_{j \neq i} \mathbf{T}_j^c \, \bigg( \frac{p_j}{p_j \cdot p_i} - \frac{q}{q \cdot p_i} \bigg) \cdot \epsilon^*(p_i,\sigma) \ket{M^{(0)}(\{ p_i \})} \; , \\[.2cm]
    \mathbf{P}_g(\sigma,c)&\, \ket{C^{(0)}_{g,i}(\{p_i\},q)} = \label{eq:Cgi} \\[.2cm]
    &- \sum_{j \neq i} \mathbf{T}_j^c \otimes \bigg( \frac{p_{i\mu} \epsilon_\nu^*(p_i,\sigma)}{p_j \cdot p_i} \big( p_j^\mu \partial_i^\nu - p_j^\nu \partial_i^\mu + iJ_j^{\mu\nu} - i\mathbf{K}_j^{\mu\nu} \big) + \frac{q_\mu \epsilon_\nu^*(p_i,\sigma))}{q \cdot p_i} \, i\mathbf{K}_i^{\mu\nu} \bigg) \ket{M^{(0)}(\{ p_i \})} \; . \nonumber
\end{align}
\begin{figure}
    \centering
    \includegraphics[width=0.9\textwidth]{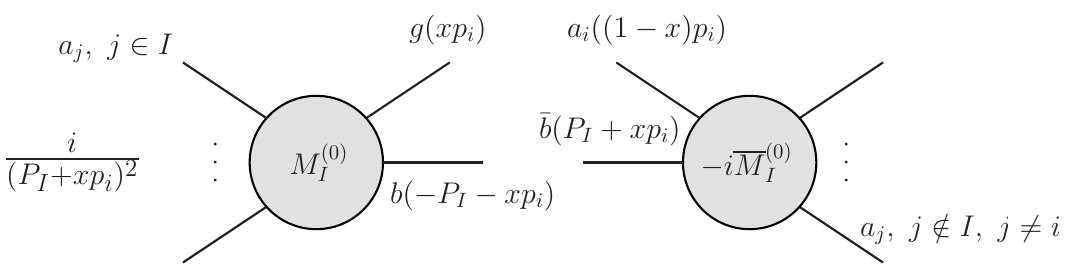}
    \caption{Class of diagrams that yields a residue contribution to the collinear-gluon amplitude. Detailed description in text following Eq.~\eqref{eq:RgiI}.}
    \label{fig:RgiI}
\end{figure}

The {\it residue} contributions, $\ket*{R^{(0)}_{g,i,I}(\{p_i\})}$, correspond to poles\footnote{If massive colour-neutral particles, e.g.\ electroweak gauge bosons, were included in the theory then the value of $x_I$ would have to be modified to include the mass of the intermediate particle.} of internal propagators that carry momentum $P_I + x \, p_i$ in the first term on the r.h.s.\ of Eq.~\eqref{eq:DeltaMgDirect} as illustrated in Fig.~\ref{fig:RgiI}:
\begin{multline} \label{eq:RgiI}
    \ip{c_1,\dots,c_{n+1};\sigma_1,\dots,\sigma_{n+1}}{R^{(0)}_{g,i,I}(\{p_i\})} = \\[.2cm] \big( 1 - x_I \big)^{-\dim(a_i)} \, \frac{1}{2p_i \cdot P_I} \sum_{\sigma c} \, M^{(0)}_I(\{p_i\},\{\sigma_i\},\{c_i\},\sigma,c) \,\widebar{M}^{(0)}_I(\{p_i\},\{\sigma_i\},\{c_i\},\sigma,c) \; .
\end{multline}
$M^{(0)}_I(\{p_i\},\{\sigma_i\},\{c_i\},\sigma,c)$ and $\widebar{M}^{(0)}_{I}(\{p_i\},\{\sigma_i\},\{c_i\},\sigma,c)$ are the tree-level amplitudes for the respective processes:
\begin{align} \label{eq:MIDef}
    &0 \to \sum_{j \in I} a_j(p_j,\sigma_j,c_j) + g(x_I \, p_i,\sigma_{n+1},c_{n+1}) + b(- P_I - x_I \, p_i,\sigma,c) \qquad \text{and} \\[.2cm] \label{eq:MIbarDef}
     &0 \to \sum_{\substack{j \not\in I \\ j \neq i}} a_j(p_j,\sigma_j,c_j) + a_i((1-x_I) \, p_i,\sigma_i,c_i) + \bar{b}(P_I + x_I \, p_i,-\sigma,c) \; ,
\end{align}
where parton $b$ is determined by flavour conservation, while $\bar{b}$ is its anti-particle. If the flavour constraint cannot be met, then the contribution for the given $I$ vanishes by definition.

The {\it anti-soft-pole} contribution, $\ket*{\bar{S}^{(0)}_{g,i}(\{p_i\},q)}$, is given by:
\begin{equation} \label{eq:Sbargi}
    \ket{\bar{S}^{(0)}_{g,i}(\{p_i\},q)} = \mathbf{E}_{i,n+1}
    \begin{cases} \sum_{j \neq i} \mathbf{Split}^{(0)}_{j,n+1 \, \leftarrow \, j}(p_j,p_i,p_j) \, \ket{M^{(0)}(\{p_i\}) \, \Big|\substack{a_i \, \to \, g \;\; \\ a_j \, \to \, \tilde{a}_j}} & \text{for $a_i \in \{q,\bar{q}\}$} \; , \\[.2cm]
    \ket{S^{(0)}_{g,i}(\{p_i\},q)} & \text{for $a_i = g$} \; ,
    \end{cases}
\end{equation}
where the splitting operator corresponds to the transition $a_ja_i \leftarrow \tilde{a}_j$. The result for $a_i \in \{q,\bar{q}\}$ is given by \eqref{eq:RgiI} in the special case $|I| = n-2$ where:
\begin{equation}
    \lim_{x_I \to 1} \ip{\dots,c_i',c_j';\dots,\sigma_i',\sigma_j'}{M^{(0)}_{I_j}(\{p_i\})} = \ip{\dots,c_i',\dots,c_j',\dots;\dots,\sigma_i',\dots,\sigma_j',\dots}{M^{(0)}(\{p_i\}) \, \Big|\substack{a_i \, \to \, g \;\; \\ a_j \, \to \, \tilde{a}_j}} \; ,
\end{equation}
\begin{multline} \label{eq:SbarSplitting}
    \lim_{x_I \to 1} \frac{\big( 1 - x_I \big)^{1-\dim(a_i)}}{\big( - P_{I_j} - x_I p_i \big)^2} \ip{c_j,c_i,c_j';\sigma_j,\sigma_i,\sigma_j'}{\widebar{M}^{(0)}_{I_j}(\{p_i\})} = \\[.2cm]
    \lim_{x_I \to 1} \big( 1 - x_I \big)^{1-\dim(a_i)} \mel{c_j,c_i;\sigma_j,\sigma_i}{\mathbf{Split}^{(0)}_{a_j a_i \, \leftarrow \, \tilde{a}_j}(p_j,(1-x_I) \, p_i,p_j)}{c_j',\sigma_j'} = \\[.2cm]
    \mel{c_j,c_i;\sigma_j,\sigma_i}{\mathbf{Split}^{(0)}_{a_j a_i \, \leftarrow \, \tilde{a}_j}(p_j,p_i,p_j)}{c_j',\sigma_j'} \; ,
\end{multline}
with:
\begin{equation}
    \qquad I_j \equiv \{ 1,\dots,n \} \setminus \{i,j\} \; , \qquad P_{I_j} = -p_i-p_j \; , \qquad \tilde{a}_j \equiv b \; .
\end{equation}
In principle, the result for $a_i = g$ can be obtained with the above method as well. However, the second equality in \eqref{eq:SbarSplitting} does not apply for $a_j = \tilde{a}_j = g$:
\begin{multline}
    \lim_{x_I \to 1} \big( 1 - x_I \big) \mel{c_j,c_i;\sigma_j,\sigma_i}{\mathbf{Split}^{(0)}_{gg \, \leftarrow \, g}(p_j,(1-x_I) \, p_i,p_j)}{c_j',\sigma_j'} \neq \\[.2cm]
    \mel{c_j,c_i;\sigma_j,\sigma_i}{\mathbf{Split}^{(0)}_{gg \, \leftarrow \, g}(p_j,p_i,p_j)}{c_j',\sigma_j'} \; .
\end{multline}
Instead, the three splitting operators \eqref{eq:SplitQGQ}, \eqref{eq:SplitQbarGQbar} and \eqref{eq:SplitGGG} yield eikonal factors. Moreover, in order to obtain the complete anti-soft pole contribution, it is still necessary to include the contribution of the last term in Eq.~\eqref{eq:HgDefGluon}. These difficulties may be overcome by using the symmetry of the collinear-gluon amplitude w.r.t.\ the exchange of the gluons $i$ and $n+1$, which straightforwardly yields \eqref{eq:Sbargi}.

Finally, the {\it linear} contribution, $\ket*{L^{(0)}_{g,i}(\{p_i\},q)}$, vanishes for $a_i \in \{ q, \bar{q} \}$, while for $a_i = g$ it is again determined by the symmetry of the collinear-gluon amplitude w.r.t.\ the exchange of the gluons $i$ and $n+1$:
\begin{equation} \label{eq:Lgi}
\begin{split}
    \ket{L^{(0)}_{g,i}(\{p_i\},q)} &= \ket{\bar{S}^{(0)}_{g,i}(\{p_i\},q)} - \ket{S^{(0)}_{g,i}(\{p_i\},q)} + \ket{\bar{C}^{(0)}_{g,i}(\{p_i\},q)} - \ket{C^{(0)}_{g,i}(\{p_i\},q)} \\[.2cm]
    &+ \frac{1}{2} \sum_I \bigg( \frac{1}{x_I} + \frac{1}{1 - x_I} \bigg) \Big( \ket{R^{(0)}_{g,i,I}(\{p_i\})} - \ket{\bar{R}^{(0)}_{g,i,I}(\{p_i\})} \Big) \; ,
\end{split}
\end{equation}
where:
\begin{equation}
    \ket{\bar{C}^{(0)}_{g,i}(\{p_i\},q)} = \mathbf{E}_{i,n+1} \, \ket{C^{(0)}_{g,i}(\{p_i\},q)} \; , \qquad
    \ket{\bar{R}^{(0)}_{g,i,I}(\{p_i\},q)} = \mathbf{E}_{i,n+1} \, \ket{R^{(0)}_{g,i,I}(\{p_i\},q)} \; .
\end{equation}

The $x$-dependence of the collinear-quark amplitude is given by:
\begin{equation} \label{eq:HqDecomposition}
\begin{split}
    \ket{H^{(0)}_{\bar{q},i}(x,\{p_i\},q)} &= \frac{1}{x} \ket{S^{(0)}_{\bar{q},i}(\{p_i\})} + \ket{C^{(0)}_{\bar{q},i}(\{p_i\},q)} + \frac{x}{1-x} \ket{\bar{S}^{(0)}_{\bar{q},i}(\{p_i\})} \\[.2cm]
    &\quad + \sum_I \bigg( \frac{1}{x_I - x} - \frac{1}{x_I} \bigg) \ket{R^{(0)}_{\bar{q},i,I}(\{p_i\})} \; .
\end{split}
\end{equation}
The soft-pole and anti-soft pole contributions are given by a similar expression to \eqref{eq:Sbargi} for the case $a_i \in \{q,\bar{q}\}$:
\begin{align} \label{eq:Sqi}
    \ket{S^{(0)}_{\bar{q},i}(\{p_i\})} &= \qquad\quad \sum_{j \neq i} \mathbf{Split}^{(0)}_{j,n+1 \, \leftarrow \, j}(p_j,p_i,p_j) \, \ket{M^{(0)}(\{p_i\}) \, \Big|\substack{a_i \, \to \, q \;\; \\ a_j \, \to \, \tilde{a}_j}} \; , \\[.2cm] \label{eq:Sbarqi}
    \ket{\bar{S}^{(0)}_{\bar{q},i}(\{p_i\})} &= \mathbf{E}_{i,n+1} \sum_{j \neq i} \mathbf{Split}^{(0)}_{j,n+1 \, \leftarrow \, j}(p_j,p_i,p_j) \, \ket{M^{(0)}(\{p_i\}) \, \Big|\substack{a_i \, \to \, \bar{q} \;\; \\ a_j \, \to \, \tilde{a}_j}} \; .
\end{align}
The splitting operator in Eq.~\eqref{eq:Sqi} corresponds to the transition $a_j\bar{q} \leftarrow \tilde{a}_j$, while that in Eq.~\eqref{eq:Sbarqi} to $a_jq \leftarrow \tilde{a}_j$. The constant contribution, $\ket*{C^{(0)}_{\bar{q},i}(\{p_i\},q)}$, corresponds to the subleading term of the soft-anti-quark expansion of the collinear-quark amplitude. An expression for this term analogous to the LBK theorem is not yet known. Hence, it has to be evaluated by using the direct expression Eq.~\eqref{eq:HqDirect} at a single convenient point. The residue contributions are obtained in analogy to Eq.~\eqref{eq:RgiI}:
\begin{multline} \label{eq:RqiI}
    \ip{c_1,\dots,c_{n+1};\sigma_1,\dots,\sigma_{n+1}}{R^{(0)}_{\bar{q},i,I}(\{p_i\})} = \\[.2cm] \big( x_I ( 1 - x_I ) \big)^{-\sfrac{1}{2}} \, \frac{1}{2p_i \cdot P_I} \sum_{\sigma c} \, M^{(0)}_I(\{p_i\},\{\sigma_i\},\{c_i\},\sigma,c) \,\widebar{M}^{(0)}_I(\{p_i\},\{\sigma_i\},\{c_i\},\sigma,c) \; .
\end{multline}
$M^{(0)}_I(\{p_i\},\{\sigma_i\},\{c_i\},\sigma,c)$ and $\widebar{M}^{(0)}_{I}(\{p_i\},\{\sigma_i\},\{c_i\},\sigma,c)$ are now the tree-level amplitudes for the respective processes:
\begin{equation}
\begin{aligned}
    &0 \to \sum_{j \in I} a_j(p_j,\sigma_j,c_j) + \bar{q}(x_I \, p_i,\sigma_{n+1},c_{n+1}) + b(- P_I - x_I \, p_i,\sigma,c) \qquad \text{and} \\[.2cm]
    &0 \to \sum_{\substack{j \not\in I \\ j \neq i}} a_j(p_j,\sigma_j,c_j) + q((1-x_I) \, p_i,\sigma_i,c_i) + \bar{b}(P_I + x_I \, p_i,-\sigma,c) \; .
\end{aligned}
\end{equation}

\subsection{Collinear convolutions} \label{sec:convolutions}

The convolution of the jet operator with the collinear-gluon amplitude can be evaluated explicitly using Eqs.~\eqref{eq:Jet}, \eqref{eq:KforJet} and \eqref{eq:HgDecomposition}:
\begin{align} \label{eq:JetConvolution}
&\mathbf{P}_g(\sigma, c) \int_0^1 \dd{x} \mathbf{J}_i^{(1)}(x, p_i, q) \ket{H_{g,i}^{(0)} (x, \{ p_i \} , q)} \nonumber\\ 
&= \frac{r_\Gamma}{\epsilon (1 - \epsilon) (1 - 2 \epsilon)} \left(- \frac{\mu^2}{s_{iq}} \right)^\epsilon  \epsilon^* (q, p_i, \sigma ) \cdot \epsilon (p_i, -\sigma)  \sum_{ c^\prime} \mathbf{P}_g(-\sigma, c^\prime) \nonumber\\ 
& \, \bigg \lbrace \mathbf{T}_i^{c^\prime} \mathbf{T}_i^c    \bigg [ -\frac{1 - 2 \epsilon}{1 + \epsilon} \left(1 - 3 \epsilon + (1 + \epsilon) \mathbf{\Sigma}_{g,i} \right) \ket{S_{g,i}^{(0)}} + (1 - 3 \epsilon - (1 - \epsilon) \mathbf{\Sigma}_{g,i} ) \ket{\bar{S}_{g,i}^{(0)}} \nonumber\\ 
& \quad + (2 - 3 \epsilon + \epsilon \mathbf{\Sigma}_{g,i}) \left( \ket{C_{g,i}^{(0)}} + \text{dim}(a_i) \ket{S_{g,i}^{(0)}} \right) - \frac{\epsilon}{2} \left(3 - \mathbf{\Sigma}_{g,i} \right) \ket{L_{g,i}^{(0)}} \\ 
& \quad + \sum_I \frac{\epsilon}{2 x_I^2 (1 - x_I)} \big ( 2x_I -2 x_I \, \mathbf{\Sigma}_{g,i} - \left(2 - x_I - x_I\, \mathbf{\Sigma}_{g,i} \right) {}_2F_1(1, 1 - \epsilon, 3 - 2 \epsilon, 1/x_I) \big ) \ket{R_{g,i,I}^{(0)}} \bigg ] \nonumber\\ 
& \, + \mathbf{T}_i^c \mathbf{T}_i^{c^\prime} \bigg [ \frac{1 - \epsilon}{1 + \epsilon} \left( 3 - 3 \epsilon  + (1 + \epsilon) \mathbf{\Sigma}_{g,i} \right) \ket{S_{g,i}^{(0)}} + \frac{\epsilon}{2} (3 - \mathbf{\Sigma}_{g,i}) \ket{\bar{S}_{g,i}^{(0)}} \nonumber\\ 
& \quad - \frac{1}{2}(4 - 3 \epsilon + \epsilon \, \mathbf{\Sigma}_{g,i}) \left( \ket{C_{g,i}^{(0)}} + \text{dim}(a_i) \ket{S_{g,i}^{(0)}} \right) + \frac{\epsilon}{2 (3 - 2 \epsilon)} (5 - 3 \epsilon - (1 - \epsilon) \mathbf{\Sigma}_{g,i}) \ket{L_{g,i}^{(0)}} \nonumber\\ 
& \quad + \sum_I \frac{\epsilon}{2 x_I^2} \big ( x_I + x_I \, \mathbf{\Sigma}_{g,i} + \left(2 - x_I - x_I \, \mathbf{\Sigma}_{g,i} \right) {}_2F_1(1, 1 - \epsilon, 3 - 2 \epsilon, 1/x_I) \big ) \ket{R_{g,i,I}^{(0)}} \bigg ] \bigg \rbrace \; ,\nonumber
\end{align}
where:
\begin{equation} \label{eq:rGamma}
    r_\Gamma = \frac{ \Gamma^2 (1 - \epsilon)\Gamma (1 + \epsilon)}{\Gamma (1 - 2 \epsilon)} \; .
\end{equation}
After expansion in $\epsilon$, Eq.~\eqref{eq:JetConvolution} exhibits singularities, which originate from the endpoints of the integration at $x = 0$ and $x = 1$. The coefficient of the $\epsilon$-pole for (anti-)quarks and gluons is provided in Eqs.~\eqref{eq:JetPoles1}, \eqref{eq:JetPoles2} and \eqref{eq:JetPoles3} in Section~\ref{sec:poles}. Notice that while the endpoint divergences in the convolution do not present an obstacle in this context, similar divergences do pose a problem in SCET. There, the convolutions typically involve operators that have already undergone renormalisation and, as a result, are defined within the confines of four-dimensional spacetime (for a detailed discussion see for example Sec.~7 of Ref.~\cite{Beneke:2019kgv}). Consequently, convolutions in SCET often require supplementary regularisation techniques, as demonstrated in Ref.~\cite{Liu:2019oav}.

In order to approximate a finite remainder of a one-loop amplitude in the 't~Hooft-Veltman scheme with Eq.~\eqref{eq:OneLoopLBK}, it is sufficient to know the $\order{\epsilon^0}$ term of the Laurent expansion of Eq.~\eqref{eq:JetConvolution}:
\begin{align}
&\left[\mathbf{P}_g(\sigma, c) e^{\epsilon \gamma_E} \int_0^1 \dd{x} \mathbf{J}_i^{(1)}(x, p_i, q) \ket{H_{g,i}^{(0)} (x, \{ p_i \} , q)}\right]_{\order{\epsilon^0}}  \nonumber\\ 
& \, = \epsilon^* (q, p_i, \sigma ) \cdot \epsilon (p_i, -\sigma)  \sum_{ c^\prime} \mathbf{P}_g(-\sigma, c^\prime) \bigg \lbrace \mathbf{T}_i^{c^\prime} \mathbf{T}_i^c    \bigg [ \left(3 - \mathbf{\Sigma}_{g,i} - (1 + \mathbf{\Sigma}_{g,i}) \ln\! \left(- \frac{\mu^2}{s_{iq}} \right)\! \right) \ket{S_{g,i}^{(0)}} \nonumber\\ 
& \quad + \left( -2 \, \mathbf{\Sigma}_{g,i} + (1 - \mathbf{\Sigma}_{g,i} ) \ln\!\left(- \frac{\mu^2}{s_{iq}} \right)\! \right) \ket{\bar{S}_{g,i}^{(0)}} \nonumber\\ 
& \quad + \left(3 + \mathbf{\Sigma}_{g,i} + 2 \ln\! \left(-\frac{\mu^2}{s_{iq}} \right)\! \right) \left( \ket{C_{g,i}^{(0)}} + \text{dim}(a_i) \ket{S_{g,i}^{(0)}} \right) - \frac{1}{2} (3 - \mathbf{\Sigma}_{g,i}) \ket{L_{g,i}^{(0)}} \nonumber\\ 
& \quad - \sum_I \frac{1}{x_I} \left(1 + \mathbf{\Sigma}_{g,i} - (2 - x_I - x_I \, \mathbf{\Sigma}_{g,i}) \ln\!\left(1 - \frac{1}{x_I} \right)\! \right) \ket{R_{g,i,I}^{(0)}} \bigg ] \\ 
& \, + \mathbf{T}_i^c \mathbf{T}_i^{c^\prime} \bigg [ \left(2\, \mathbf{\Sigma}_{g,i} + (3 + \mathbf{\Sigma}_{g,i}) \ln\!\left(- \frac{\mu^2}{s_{iq}} \right)\! \right) \ket{S_{g,i}^{(0)}} + \frac{1}{2} (3 - \mathbf{\Sigma}_{g,i}) \ket{\bar{S}_{g,i}^{(0)}} \nonumber\\ 
& \quad - \frac{1}{2} \left(9 + \mathbf{\Sigma}_{g,i} + 4 \ln\!\left(-\frac{\mu^2}{s_{iq}} \right)\! \right) \left(\ket{C_{g,i}^{(0)}} + \text{dim}(a_i) \ket{S_{g,i}^{(0)}} \right) + \frac{1}{6} (5 - \mathbf{\Sigma}_{g,i}) \ket{L_{g,i}^{(0)}} \nonumber\\ 
& \quad + \sum_I \frac{1}{2 x_I} \left( 5 - 2 x_I + (1 - 2 x_I) \mathbf{\Sigma}_{g,i} - 2 (1 - x_I) (2 - x_I - x_I \, \mathbf{\Sigma}_{g,i}) \ln\!\left(1 - \frac{1}{x_I} \right)\!  \right) \ket{R_{g,i,I}^{(0)}} \bigg ] \bigg \rbrace \; ,\nonumber
\end{align}
where we have removed the Euler-Mascheroni constant $\gamma_E$ as would be done in the $\overline{\text{MS}}$ scheme.

The convolution of the flavour-off-diagonal jet operator with the collinear-quark amplitude can be evaluated explicitly using Eqs.~\eqref{eq:JetTilde} and \eqref{eq:HqDecomposition}:
\begin{align}
&\mathbf{P}_i (\sigma_i, c_i) \mathbf{P}_g(\sigma, c) \int_0^1 \dd{x} \mathbf{\tilde{J}}_i^{(1)} (x, p_i, q) \ket{H_{\bar{q}, i}^{(0)}(x, \{p_i \}, q) } \nonumber\\ 
&\, = \frac{r_\Gamma}{(1 - \epsilon) (1 - 2 \epsilon)} \left(- \frac{\mu^2}{s_{iq}} \right)^\epsilon \epsilon^*(q, p_i, \sigma) \cdot \epsilon^*(p_i, \sigma_i) \sum_{\sigma^\prime c^\prime} \sum_{c_i^\prime} \mathbf{P}_i (-\sigma^\prime, c_i^\prime) \mathbf{P}_{n + 1} (\sigma^\prime, c^\prime) \nonumber\\ 
&\, \, \bigg \lbrace(T_q^{c_i} T_q^{c})_{c^\prime c_i^\prime}
\bigg [ 2\sigma_i\sigma'\ket{S_{\bar{q},i}^{(0)}} + \left(\frac{1-(2-\epsilon)\sigma_i\sigma'}{\epsilon}+\frac{1}{2(3-2\epsilon)}\right)\ket{\bar{S}_{\bar{q},i}^{(0)}} + \left(\sigma_i\sigma'-\frac{1}{2(3-2\epsilon)}\right)\ket{C_{\bar{q},i}^{(0)}} \nonumber\\
&\quad + \sum_I \frac{1}{x_I}\left(2x_I^2 - (1+2x_I)\sigma_i\sigma'+\frac{1}{2(3-2\epsilon)}+x_I(1-2x_I+2\sigma_i\sigma') {}_2F_1 (1, 1 - \epsilon, 2 - 2 \epsilon, 1/x_I) \right)\ket{R_{\bar{q},i,I}^{(0)}}\bigg] \nonumber\\
&\,\,+(T_q^c T_q^{c_i})_{c^\prime c_i^\prime} \bigg [\left(2\sigma_i\sigma' - \frac{1+2\sigma_i\sigma'}{\epsilon}\right)\ket{S_{\bar{q},i}^{(0)}}
+\left(\sigma_i\sigma'-\frac{1}{2(3-2\epsilon)}\right)\ket{\bar{S}_{\bar{q},i}^{(0)}} + \left(\sigma_i\sigma'+\frac{1}{2(3-2\epsilon)}\right)\ket{C_{\bar{q},i}^{(0)}} \nonumber\\
&\quad+\sum_I\frac{1}{x_I}\Big(2x_I-2x_I^2-(1-2x_I)\sigma_i\sigma'-\frac{1}{2(3-2\epsilon)}\nonumber\\
&\qquad+(1-x_I)(1-2x_I+2\sigma_i\sigma'){}_2F_1 (1, 1 - \epsilon, 2 - 2 \epsilon, 1/x_I)\Big) \ket{R_{\bar{q},i,I}^{(0)}} \bigg]\bigg \rbrace \; .
\end{align}
The $\mathcal{O}(\epsilon^0)$ term of the Laurent expansion is given by:
\begin{align}
&\left[\mathbf{P}_i (\sigma_i, c_i) \mathbf{P}_g(\sigma, c) e^{\epsilon \gamma_E}\int_0^1 \dd{x} \mathbf{\tilde{J}}_i^{(1)} (x, p_i, q) \ket{H_{\bar{q}, i}^{(0)}(x, \{p_i \}, q) }\right]_{\mathcal{O}(\epsilon^0)} \nonumber\\ 
&\, = \epsilon^*(q, p_i, \sigma) \cdot \epsilon^*(p_i, \sigma_i) \sum_{\sigma^\prime c^\prime} \sum_{c_i^\prime} \mathbf{P}_i (-\sigma^\prime, c_i^\prime) \mathbf{P}_{n + 1} (\sigma^\prime, c^\prime) \nonumber\\ 
&\, \, \bigg \lbrace(T_q^{c_i} T_q^c)_{c^\prime c_i^\prime} \bigg [2 \sigma_i \sigma^\prime \ket{S_{\bar{q},i}^{(0)}} + \left(\frac{19}{6} - 5 \sigma_i \sigma^\prime + (1 - 2 \sigma_i \sigma^\prime) \ln\! \left(-\frac{\mu^2}{s_{iq}} \right)\! \right) \ket{\bar{S}_{\bar{q},i}^{(0)}}  - \left(\frac{1}{6} - \sigma_i \sigma^\prime \right) \ket{C_{\bar{q},i}^{(0)}} \nonumber\\ 
& \quad + \sum_I \left( \frac{1}{6 x_I} \left( 1 + 12 x_I^2 - 6 (1 + 2 x_I) \sigma_i \sigma^\prime  \right) - x_I (1 - 2 x_I + 2 \sigma_i \sigma^\prime) \ln\! \left(1 - \frac{1}{x_I} \right)\! \right) \ket{R_{\bar{q},i,I}^{(0)}} \bigg ] \nonumber\\ 
& \, \, + (T_q^c T_q^{c_i})_{c^\prime c_i^\prime} \bigg [\! - \!\left( 3 + 4 \sigma_i \sigma^\prime + (1 + 2 \sigma_i \sigma^\prime) \ln\! \left( - \frac{\mu^2}{s_{iq}} \right) \!\right) \ket{S_{\bar{q},i}^{(0)}} - \left(\frac{1}{6} - \sigma_i \sigma^\prime\right) \ket{\bar{S}_{\bar{q},i}^{(0)}} \nonumber\\
&\quad+ \left( \frac{1}{6} + \sigma_i \sigma^\prime \right) \ket{C_{\bar{q},i}^{(0)}} - \sum_I \bigg( \frac{1}{6 x_I} \left(1 - 12 x_I + 12 x_I^2 + 6 (1- 2 x_I) \sigma_i \sigma^\prime \right) \nonumber\\
&\qquad +(1 - x_I) (1 - 2 x_I + 2\sigma_i \sigma^\prime) \ln\! \left(1 - \frac{1}{x_I} \right) \!\!\bigg) \ket{R_{\bar{q},i,I}^{(0)}} \bigg ] \bigg \rbrace \; .
\end{align}

\subsection{Proof based on the expansion-by-regions method} \label{sec:proof}

Theorem \ref{sec:theorem} has been obtained by applying the expansion-by-regions method \cite{Beneke:1997zp} (see also Refs.~\cite{Smirnov:2002pj, Jantzen:2011nz}, and for its application to the subleading soft-gluon expansion, see Ref.~\cite{Bonocore:2014wua}). The method is anchored in dimensional regularisation, and can be used to expand Feynman diagrams in any parameter. There are three difficulties: 1) identification of contributing regions, 2) appearance of unregulated integrals, 3) application to a large number of diagrams. Problem 1) has been solved for several standard expansions. The soft expansion has been analysed most recently in Refs.~\cite{Gervais:2017yxv, Laenen:2020nrt, Engel:2021ccn, Engel:2023ifn} albeit for soft-photon emissions. The most important observation is the appearance of a collinear region besides the expected hard and soft regions. Although the collinear region has been anticipated already in Ref.~\cite{DelDuca:1990gz}, the latter analysis has been shown to be incomplete. Irrespective of the listed publications, the identification of contributing regions can nowadays be performed automatically with dedicated tools \cite{Jantzen:2012mw, Ananthanarayan:2018tog, Heinrich:2021dbf}. As far as problem 2) is concerned, it turns out that no unregulated integrals appear in the soft expansion considered here. Finally, problem 3) is alleviated by organising the contributions according to physical intuition.

The three contributing regions, hard, soft and collinear, are rather classes of regions defined by a scaling of the loop momentum w.r.t.\ the expansion parameter. In each class, an actual region is defined by a loop-momentum routing.
Actually, momentum routing is relevant in all but the hard region. The latter is defined by assuming that each component of the loop momentum is large compared to the expansion parameter. This region is the easiest to analyse. In fact, the respective Feynman integrands are obtained by Taylor expansion in the momentum shifts $\delta_i$ and the soft-gluon momentum $q$. It follows immediately that the hard-region contribution is given by the first term in Eq.~\eqref{eq:OneLoopLBK}. This corresponds to Eq.~\eqref{eq:LBK1} upon replacement of tree-level amplitudes by their one-loop counterparts.

The soft and collinear regions present more subtleties and are analysed below. One important property should already be stressed at this point. Each region has a different $d$-dimensional scaling w.r.t.\ to the expansion parameter. Hence, each region is gauge-invariant on its own. We will exploit this property to make the calculations as simple as possible. The only subtle point is that some gauges, e.g.\ the lightcone gauge, may generate additional singularities and hence additional regions. These unphysical regions must cancel entirely upon summation of the contributions in a given class due to the gauge invariance of the original amplitude. With the choices made below, no unphysical regions appear in the first place.

\subsubsection*{Soft regions}

\begin{figure}
\begin{minipage}[t]{0.45\textwidth}
    \centering
    \includegraphics[width=0.6\textwidth]{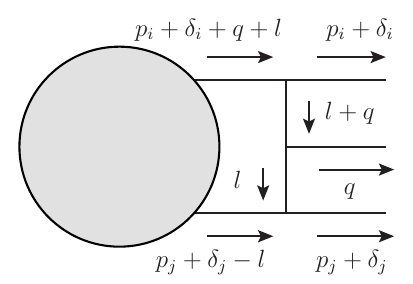}
    \caption{Routing of the loop-momentum $l$ in the $(i,j)$-soft region.}
    \label{fig:SoftRegionRouting}
\end{minipage}
\hspace{.5cm}
\begin{minipage}[t]{0.45\textwidth}
    \centering
    \raisebox{1.9mm}{
    \includegraphics[width=0.6\textwidth]{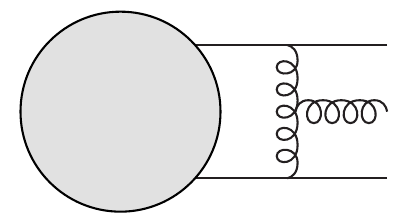}}
    \caption{Flavour-diagonal soft-region diagram. Solid lines represent an arbitrary parton, i.e.\ quark, anti-quark or gluon.}
    \label{fig:SoftRegionDiagonal}
\end{minipage}
\end{figure}
\begin{figure}
\centering
\includegraphics[scale=0.6]{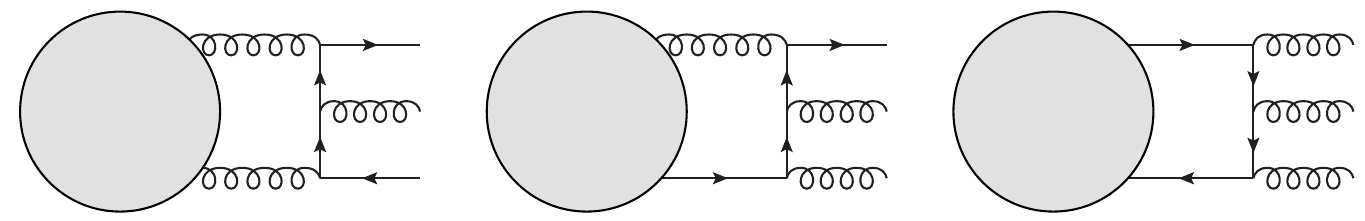}
\caption{Flavour-off-diagonal soft-region diagrams.}
\label{fig:SoftRegionOffDiagonal}
\end{figure}

In any soft region, the loop momentum, $l$, is assumed be of the order of the soft-gluon momentum, $l^\mu = \order{\lambda}$. A particular soft region is defined by selecting a pair of external partons $i,j$. We differentiate between flavour-diagonal, Fig.~\ref{fig:SoftRegionDiagonal}, and flavour-off-diagonal contributions, Fig.~\ref{fig:SoftRegionOffDiagonal}. In principle, the soft gluon may attach anywhere else on the visible lines in Figs.~\ref{fig:SoftRegionDiagonal} and \ref{fig:SoftRegionOffDiagonal}. However, a scaling argument demonstrates that the shown topologies are the only ones that yield non-vanishing integrals after expansion, since alternative topologies result in scaleless integrals.

The momentum routing in the $(i,j)$-soft region is specified in Fig.~\ref{fig:SoftRegionRouting}. The calculation is conveniently performed in the Feynman gauge. The matrix element represented by the shaded circle is expanded in $\delta_l$, $l$ and $q$ just as in Section~\ref{sec:LBKderivation}. In the case of flavour-off-diagonal diagrams, the expansion is trivial and amounts to setting these parameters to zero. Tensor integrals are reduced to scalar integrals with Passarino-Veltman reduction \cite{Passarino:1978jh}. The diagrams are expressed in terms of a single non-vanishing integral:
\begin{equation}
\begin{split}
    I^{\text{soft}} &= \mu^{2\epsilon} \int \frac{d^d l}{i \pi^{d/2}} \frac{(p_i + \delta_i) \cdot (p_j + \delta_j)}{[l^2 +i0^+ ] [(l + q)^2 + i0^+] [(p_i + \delta_i) \cdot (l + q) + i0^+] [-(p_j + \delta_j) \cdot l + i0^+]} \\[.2cm]
    &= \frac{r_{\text{Soft}}}{\epsilon^2} \frac{4s^{(\delta)}_{ij}}{s^{(\delta)}_{iq} s^{(\delta)}_{jq}} \Bigg(- \frac{\mu^2 s^{(\delta)}_{ij}}{s^{(\delta)}_{iq} s^{(\delta)}_{jq}} \Bigg)^\epsilon \; ,
\end{split}
\end{equation}
where we have not yet expanded in $\delta_i$, $\delta_j$. $r_{\text{Soft}}$ has been defined in \eqref{eq:rSoft} while the invariants $s^{(\delta)}_{\dots}$ in \eqref{eq:invariantsDelta}. The results are summarized in Eqs.~\eqref{eq:Soft} and \eqref{eq:SoftTilde}. They have all the desired properties: they satsify the Ward identity w.r.t.\ to the soft-gluon momentum, they are expressed through gauge-invariant reduced scattering amplitudes, the occurring differential operators are consistent with on-shellness and momentum conservation. As expected, each of these properties applies in a single $(i,j)$-soft region. Notice, however, that momentum conservation requires symmetrisation w.r.t.\ $i$ and $j$ due to the fact that Eq.~\eqref{eq:Soft} is written in a non-symmetric form.

\subsubsection*{Collinear regions}

\begin{figure}
\centering
\includegraphics[width=0.8\textwidth]{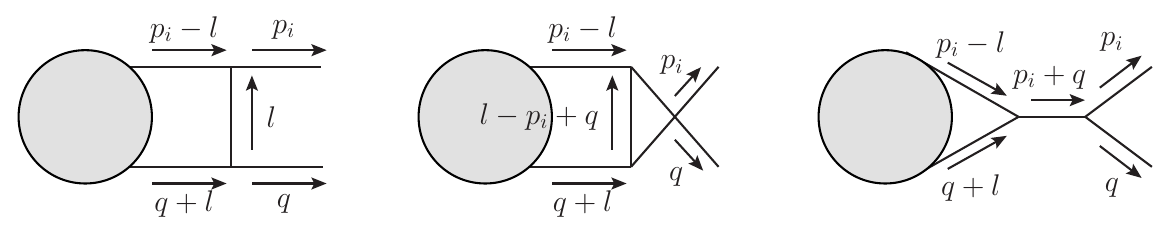}
\caption{Routing of the loop-momentum $l$ in the three topologies occurring in the $i$-collinear region.}
\label{fig:CollinearRegionRouting}
\end{figure}

A particular collinear region is defined by selecting a parton $i$ whose momentum specifies the {\it collinear direction} $n$ with $n \, \propto \, p_i$. An {\it anti-collinear direction} $\bar{n}$, $\bar{n}^2 = 0$, $\bar{n} \, \cancel\propto \, n$ must also be specified. In principle, the only natural choice is $\bar{n} \, \propto \, q$. In the following, we will nevertheless keep $\bar{n}$ generic albeit normalised to conveniently satisfy $n \cdot \bar{n} = \sfrac{1}{2}$. An arbitrary vector $k$ can now be decomposed as follows:
\begin{equation}
    k = k_+ n + k_- \bar{n} + k_\perp \; , \qquad k_\pm \in \mathbb{R} \; , \qquad k_\perp \cdot n = k_\perp \cdot \bar{n} = 0 \; , \qquad k_\perp^2 \leq 0 \; , \qquad k^2 = k_+ k_- + k_\perp^2 \; .
\end{equation}

The expanded amplitude will be calculated in the lightcone gauge with gauge vector $\bar{n}$. The use of a physical gauge simplifies the analysis of the singularity structure of diagrams and is particularly important in the study of collinear radiation. In particular, our gauge choice yields results that do not necessitate derivatives of process-dependent scattering amplitudes. This is at variance with Ref.~\cite{Laenen:2020nrt}, where tests of factorisation formulae for soft-photon radiation were performed in the Feynman gauge, which led to the appearance of different jet operators than ours. Finally, the disappearance of $\bar{n}$ from the final expressions will serve as a test of independence from the particular physical gauge chosen.

The routing of the loop momentum $l$ is specified in Fig.~\ref{fig:CollinearRegionRouting} for the three topologies characteristic of the $i$-collinear region. The integration measure is given by:
\begin{equation}
    d^d l = \frac{1}{2} \dd{l_+} \dd{l_-} \dd[d-2]{l_\perp} \; .
\end{equation}
Expansion in $\lambda$ is performed according to:
\begin{equation} \label{eq:CollinearScaling}
    l_+ = \order{1} \; , \qquad l_\perp = \order{\lambda^{\sfrac{1}{2}}} \; , \qquad l_- = \order{\lambda} \; .
\end{equation}
Propagator denominators are, therefore, approximated as follows:
\begin{equation}
\begin{gathered}
    (l+q)^2 + i0^+ \approx l_+ (l_- + q_-) + l_\perp^2 + i0^+ \; , \qquad
    (l-p_i)^2 + i0^+ \approx (l_+ - p_{i+}) l_- + l_\perp^2 + i0^+ \; , \\[.2cm]
    (l-p_i+q)^2 + i0^+ \approx (l_+ - p_{i+}) (l_- + q_-) + l_\perp^2 + i0^+ \; .
\end{gathered}
\end{equation}
Expansion of the actual propagators generates, of course, further terms polynomial in $q_-$, $l_-$ and $l_\perp$ accompanied by higher powers of the propagator denominators. The part of the integrand represented by the shaded circle in Fig.~\ref{fig:CollinearRegionRouting} must also be expanded according to \eqref{eq:CollinearScaling}. Hence, this part depends non-trivially on $l_+$, while any dependence on $l_-$ and $l_\perp$ is introduced through differential operators $( l_- \pdv*{l_-} )^{k_1} ( l_\perp \cdot \pdv*{l_\perp} )^{k_2}$ with the derivatives evaluated at vanishing $l_-$ and $l_\perp$. One can factor out $p_{i+}$ and $q_-$ from the integrand term-by-term. This is achieved by the change of variable:
\begin{equation}
    l_+ \equiv x \, p_{i+} \; ,
\end{equation}
and the rescalings $l_- \to q_- \, l_-$, $l_\perp^2 \to p_{i+} \, q_- \, l_\perp^2$. In consequence, expanded integrals are proportional to $\big( p_{i+} q_- \big)^{-\epsilon} = \big( 2 p_i \cdot q \big)^{-\epsilon}$. Furthermore, both $p_{i+}$ and $q_-$ must be present in the propagator denominators without possibility to remove them by loop-momentum shifts, or otherwise a given integral is scaleless. 

\begin{figure}
\centering
\includegraphics[scale=0.5]{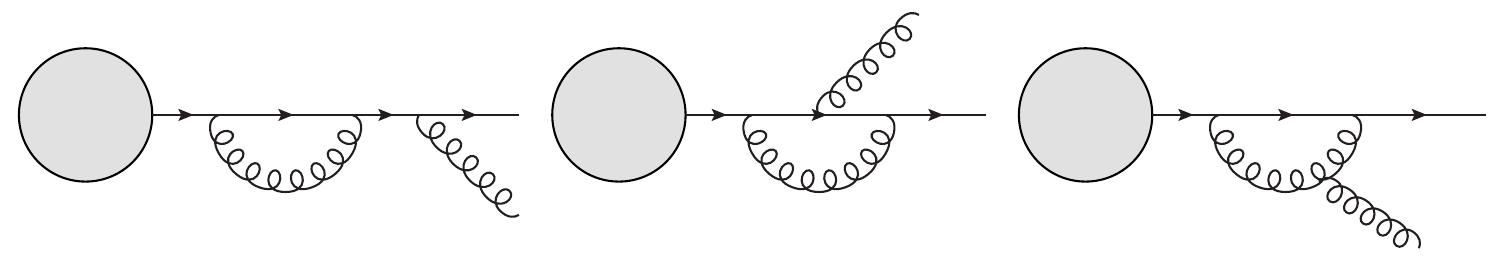}
\caption{Collinear-region diagrams with soft-gluon emission from an external outgoing quark line.}
\label{fig:SingleJetQuark}
\end{figure}

\begin{figure}
\centering
\includegraphics[scale=0.5]{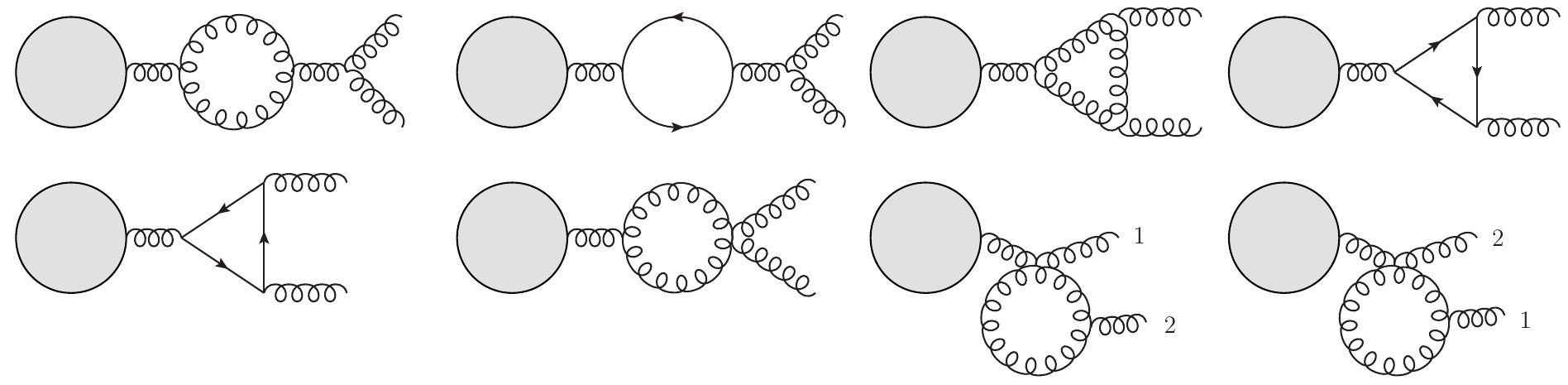}
\caption{Collinear-region diagrams with soft-gluon emission from an external outgoing gluon line. In the last two diagrams, the soft-gluon is indicated by the label ``1'', while the hard gluon by the label ``2''. These two diagrams do not contribute, however, since the external momenta are on-shell and the resulting integrals are scaleless.}
\label{fig:SingleJetGluon}
\end{figure}

After expansion, integration over $l_-$ can be performed by closing the integration contour in the upper complex half-plane, and taking residues at:
\begin{equation} \label{eq:residues}
    \frac{l_\perp^2 + i0^+}{-l_+} \; , \qquad  - q_- + \frac{l_\perp^2 + i0^+}{-l_+} \; , \qquad \frac{l_\perp^2 + i0^+}{p_+ - l_+} \; , \qquad - q_- + \frac{l_\perp^2 + i0^+}{p_+ - l_+} \; .
\end{equation}
The first two of the residues contribute only for $l_+ < 0$, while the second two only for $l_+ < p_+$. The final integration over $l_\perp$  effectively only involves $(d-2)$-dimensional massive vacuum integrals. For this reason, any contribution odd in $l_\perp$ vanishes.

In the case of collinear-region contributions depicted in Figs.~\ref{fig:SingleJetQuark} and \ref{fig:SingleJetGluon} the loop-momentum integration can be performed explicitly. In particular, denoting by $\sigma$,$c$ and $\sigma_i$,$c_i$ the helicity and colour of the soft-gluon and parton $i$ respectively, one finds:
\begin{multline} \label{eq:SingleJetQuark}
    \text{Fig.~\ref{fig:SingleJetQuark}} = r_\Gamma \bigg( -\frac{\mu^2}{s_{iq}} \bigg)^{\epsilon} \mathbf{P}_i(\sigma_i,c_i) \mathbf{T}_i^c \, \epsilon^*_\mu(q,p_i,\sigma) \bar{u}(p_i,\sigma_i) \bigg[ \frac{C_F-C_A}{1-2\epsilon} \frac{\gamma^\mu\slashed{q}}{2p_i\cdot q} \\[.2cm]
    - \frac{1}{1-\epsilon} \left( \frac{2C_F}{1-2\epsilon}+\frac{C_A}{\epsilon} \right) \frac{\gamma^\mu\slashed{\bar{n}}}{2p_i\cdot \bar{n}} - \frac{2}{1-\epsilon} \left(\frac{C_F}{\epsilon}-\frac{C_A}{1+\epsilon}\right) \frac{\bar{n}^\mu}{p_i\cdot \bar{n}} \bigg] \pdv{\bar{u}(p_i,\sigma_i)} \ket{M^{(0)}} \; ,
\end{multline}
\begin{multline} \label{eq:SingleJetGluon}
    \text{Fig.~\ref{fig:SingleJetGluon}} = r_\Gamma \bigg( -\frac{\mu^2}{s_{iq}} \bigg)^{\epsilon} \mathbf{P}_i(\sigma_i,c_i) \mathbf{T}_i^c \, \epsilon^*_\mu(q,p_i,\sigma) \epsilon_\beta^*(p_i,\sigma_i) \\[.2cm]
    \times \bigg\{ - C_A \bigg[ \frac{1}{1-2\epsilon} \left( \frac{1}{3-2\epsilon}\frac{g^{\mu\beta} q^\alpha}{p_i\cdot q} + \frac{1}{(1-\epsilon)\epsilon}\frac{g^{\mu\beta} \bar{n}^\alpha}{p_i\cdot \bar{n}} \right)
    + \frac{2}{(1-\epsilon)(1+\epsilon)\epsilon} \frac{\bar{n}^\mu g^{\beta\alpha}}{p_i\cdot \bar{n}} \bigg] \\[.2cm]
    + T_F n_l \frac{2}{(1-\epsilon)(1-2\epsilon)(3-2\epsilon)} \frac{g^{\mu\beta} q^\alpha}{p_i\cdot q} \bigg\} \pdv{\epsilon^*_\alpha(p_i,\sigma_i)} \ket{M^{(0)}} \; ,
\end{multline}
with $r_\Gamma$ defined in \eqref{eq:rGamma}.

\begin{figure}
\centering
\includegraphics[scale=0.5]{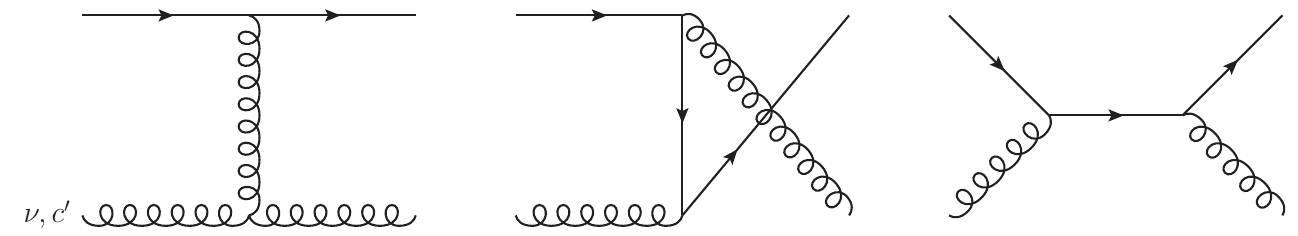}
\caption{Subdiagrams contributing to the jet operator for an outgoing quark. Lines on the left-hand sides of the diagrams are not amputated and are represented by propagators in the integrand. Integration over $l_-$, $l_\perp$ is included in the expressions for the diagrams.}
\label{fig:JetOperatorQuark}
\end{figure}
\begin{figure}
\centering
\includegraphics[scale=.5]{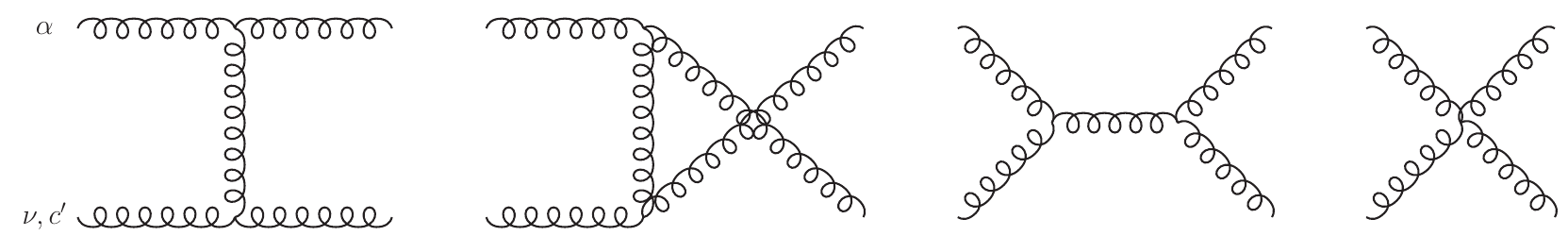}
\caption{Subdiagrams contributing to the jet operator for a gluon. Description as in Fig.~\ref{fig:JetOperatorQuark}. A factor of $\sfrac{1}{2}$ must be included in the calculation of the diagrams in order to compensate for the symmetry of the amplitude represented by the shaded circle in Fig.~\ref{fig:CollinearRegionRouting}.}
\label{fig:JetOperatorGluon}
\end{figure}
\begin{figure}
\centering
\includegraphics[scale=.5]{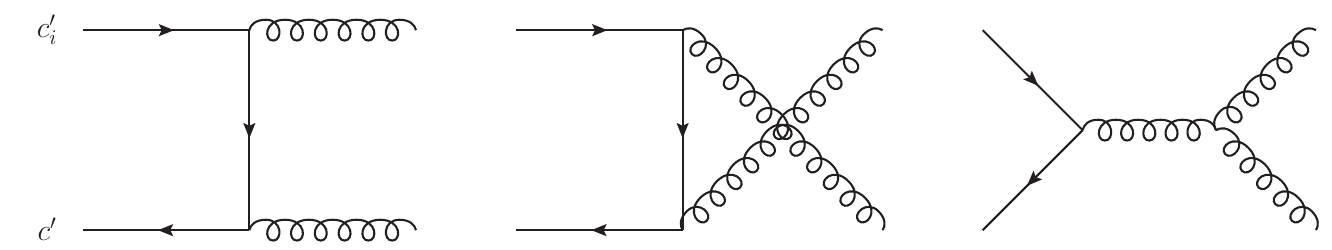}
\caption{Subdiagrams contributing to the flavour-off-diagonal jet operator. Description as in Fig.~\ref{fig:JetOperatorQuark}.}
\label{fig:JetOperatorTilde}
\end{figure}

The remaining collinear-region contributions require the knowledge of the $x$-dependence of the part of the integrand represented by the shaded circle in Fig.~\ref{fig:CollinearRegionRouting}. It turns out that no derivatives in $l_-$, $l_\perp$ are needed at $\order{\lambda^0}$, since contributions containing differential operators $( l_- \pdv*{l_-} )^{k_1} ( l_\perp \cdot \pdv*{l_\perp} )^{k_2}$, $2k_1+k_2 \leq 2$ cancel. Hence, integration over $l_-$, $l_\perp$  only involves the subdiagrams  depicted in Figs.~\ref{fig:JetOperatorQuark}, \ref{fig:JetOperatorGluon} and \ref{fig:JetOperatorTilde}. The results are as follows:
\begin{multline} \label{eq:QuarkJet}
    \text{Fig.~\ref{fig:JetOperatorQuark}} \equiv \mathbf{J}^{\nu,c'}_{q} = \frac{\Gamma(1+\epsilon)}{1-\epsilon} \bigg(- \frac{\mu^2}{s_{iq}} \bigg)^{\epsilon} (x(1-x))^{-\epsilon} \, \mathbf{P}_i(\sigma_i,c_i) \left( \mathbf{T}_i^{c} \mathbf{T}_i^{c'} + \frac{1}{x} i f^{cdc'} \mathbf{T}_i^d \right) \\[.2cm]
    \times \epsilon^*_\mu(q,p_i,\sigma) \, \bar{u}(p_i,\sigma_i) \big( 2 g^{\mu\beta} - x \, \gamma^\mu \gamma^\beta \big) g_{\perp\beta}{}^\nu \; ,
\end{multline}
\begin{multline} \label{eq:GluonJet}
    \text{Fig.~\ref{fig:JetOperatorGluon}} \equiv \mathbf{J}^{\alpha\nu,c'}_{g} = \frac{\Gamma(1+\epsilon)}{1-\epsilon} \bigg(- \frac{\mu^2}{s_{iq}} \bigg)^{\epsilon}(x(1-x))^{-\epsilon} \, \mathbf{P}_i(\sigma_i,c_i) \left( \mathbf{T}_i^{c} \mathbf{T}_i^{c'} + \frac{1}{x} i f^{cdc'} \mathbf{T}_i^d \right) \\[.2cm]
    \times \epsilon^*_\mu(q,p_i,\sigma) \, \epsilon^*_\beta(p_i,\sigma_i) \, \bigg( g_\perp^{\mu\nu} g_\perp^{\beta\alpha} - x \, g^{\mu\beta} g_\perp^{\nu\alpha} + \frac{x}{1-x} \, g_\perp^{\mu\alpha} g_\perp^{\beta\nu} \bigg) \; ,
\end{multline}
\begin{multline} \label{eq:TildeJet}
    \text{Fig.~\ref{fig:JetOperatorTilde}} \equiv \tilde{J}_{c'c_i'} = \frac{\Gamma(1 + \epsilon)}{1 - \epsilon} \bigg(- \frac{\mu^2}{s_{iq}} \bigg)^{\epsilon}(x(1-x))^{-\epsilon} \left( T^c T^{c_i} + i x f^{cdc_i} T^d \right)_{c' c_i^\prime} \\[.2cm] 
 \times \epsilon^*_\mu(q,p_i,\sigma) \, \epsilon^*_\beta(p_i,\sigma_i) \, \slashed{p}_i \big( 2 x \, g^{\mu\beta} - \gamma^\mu \gamma^\beta \big) \; ,
\end{multline}
where:
\begin{equation}
    g_\perp^{\mu\nu} \equiv g^{\mu\nu} - \frac{p_i^\mu \bar{n}^\nu + p_i^\nu \bar{n}^\mu}{p_i \cdot \bar{n}} \; .
\end{equation}
The contributions of the residues in $l_-$ at the points listed in \eqref{eq:residues} conspire to cancel unless:
\begin{equation}
    x \in [0,1] \; .
\end{equation}

Since Figs.~\ref{fig:SingleJetQuark} and \ref{fig:SingleJetGluon} have the structure of Fig.~\ref{fig:CollinearRegionRouting}, one might expect that the results presented in Eqs.~\eqref{eq:SingleJetQuark} and \eqref{eq:SingleJetGluon} can be obtained by integrating $\mathbf{J}^{\nu,c'}_{q}$, $\mathbf{J}^{\alpha\nu,c'}_{g}$ and $\tilde{J}_{c'c_i'}$ with appropriate functions of $x$. This is indeed the case:
\begin{align} \label{eq:SingleJetQuarkAlt}
     &\text{Fig.~\ref{fig:SingleJetQuark}} = \int_0^1 \dd{x} \mathbf{J}^{\nu,c'}_{q} \, \mathbf{T}_i^{c'} \frac{1}{p_i \cdot q} \left( - \frac{1}{2} \gamma_\nu \slashed{q} - \frac{1}{x} q_\nu \right) \pdv{\bar{u}(p_i,\sigma_i)} \ket{M^{(0)}} \; , \\[.2cm] \label{eq:SingleJetGluonAlt}
&\begin{multlined}
    \text{Fig.~\ref{fig:SingleJetGluon}} = \int_0^1 \dd{x} \mathbf{J}^{\alpha\nu,c'}_{g} \, \mathbf{T}_i^{c'} \frac{1}{p_i \cdot q} \left( -(1-2x) g_{\alpha\nu} q_\beta - \frac{q_\nu g_{\alpha\beta}}{x} + \frac{q_\alpha g_{\nu\beta}}{1-x} \right) \pdv{\epsilon_\beta^*(p_i,\sigma_i)} \ket{M^{(0)}} \\[.2cm]
    + n_l \int_0^1 \dd{x} \Tr[ \tilde{J}_{c'c_i'} \frac{\slashed{q}}{p_i \cdot q} ] T^{c_i''}_{c_i'c'} \frac{q}{p_i \cdot q} \cdot \pdv{\epsilon^*(p_i,\sigma_i'')} \mathbf{P}_i(\sigma_i'',c_i'') \ket{M^{(0)}} \; .
\end{multlined}
\end{align}
The choice of the helicity $\sigma_i''$ in the contribution proportional to $n_l$ in Eq.~\eqref{eq:SingleJetGluonAlt} does not affect the result.

The relevance of Eqs.~\eqref{eq:SingleJetQuarkAlt} and \eqref{eq:SingleJetGluonAlt} becomes apparent after consultation of the expressions for the collinear-gluon and collinear-quark amplitudes, \eqref{eq:HgDefQuark}, \eqref{eq:HgDefGluon}, \eqref{eq:DeltaMgDirect} and \eqref{eq:HqDirect}. Clearly, soft-gluon emissions from external lines are correctly accounted for by the convolutions of either $\mathbf{J}^{\nu,c'}_{q}$  with $\ket*{H^{(0)}_{g,i}}$, or of $\mathbf{J}^{\alpha\nu,c'}_{g}$ with $\ket*{H^{(0)}_{g,i}}$ and $\tilde{J}_{c'c_i'}$ with $\ket*{H^{(0)}_{\bar{q},i}}$. In both cases, it is still necessary to remove the external wave functions of partons $i$ and $n+1$ from the collinear amplitudes. The convolutions thus provide the entirety of the contribution of the $i$-collinear region.

At this point we recall what has been proven in Section~\ref{sec:CollinearAmplitudes}, namely that $\ket*{H^{(0)}_{g,i}}$ satisfies the Ward identity w.r.t.\ any gluon. Hence, terms proportional to $p_i^\nu$ in Eqs.~\eqref{eq:QuarkJet}, \eqref{eq:GluonJet} and additionally to $p_i^\alpha$ in Eq.~\eqref{eq:GluonJet} vanish after contraction with the collinear-gluon amplitude. Equivalently, removing $\bar{n}$-dependent terms hidden in $g_\perp^{\mu\nu}$ from $\mathbf{J}^{\nu,c'}_{q}$ and $\mathbf{J}^{\alpha\nu,c'}_{g}$ does not affect the $i$-collinear-region contribution. In consequence, our results do not depend on the anti-collinear direction and thus the particular physical gauge used to derive them.

The result for the jet operator \eqref{eq:Jet} for $a_i = q$ now directly follows from Eqs.~\eqref{eq:QuarkJet} and \eqref{eq:Kdef}. In order to obtain \eqref{eq:Jet} for $a_i = g$, it is necessary to first transform Eq.~\eqref{eq:GluonJet} by exploiting the symmetry of the collinear-gluon amplitude w.r.t.\ gluons $i$ and $n+1$ together with the Jacobi identity in the form:
\begin{equation}
    \bigg( T^c_g T^{c'}_g + \frac{1}{x} if^{cdc'} T_g^d \bigg)_{c_ic_i'} = \bigg( \frac{1-x}{x} \, T^c_g T^{c_i'}_g + \frac{1}{x} if^{cdc_i'} T_g^d \bigg)_{c_ic'} \; .
\end{equation}
Eq.~\eqref{eq:GluonJet} is then equivalent to:
\begin{multline} \label{eq:GluonJetEquivalent}
    \frac{\Gamma(1+\epsilon)}{1-\epsilon} \bigg(- \frac{\mu^2}{s_{iq}} \bigg)^{\epsilon}(x(1-x))^{-\epsilon} \, \mathbf{P}_i(\sigma_i,c_i) \left( \mathbf{T}_i^{c} \mathbf{T}_i^{c'} + \frac{1}{x} i f^{cdc'} \mathbf{T}_i^d \right) \\[.2cm]
    \times \epsilon^*_\mu(q,p_i,\sigma) \, \epsilon^*_\beta(p_i,\sigma_i) \, \Big( (2 - x) \, g^{\mu\nu} g^{\beta\alpha} - x \, \big( g^{\mu\beta} g^{\nu\alpha} - g^{\mu\alpha} g^{\beta\nu} \big) \Big) \; ,
\end{multline}
which, together with \eqref{eq:Kdef}, indeed yields \eqref{eq:Jet}. Finally, Eq.~\eqref{eq:JetTilde} is obtained from Eq.~\eqref{eq:TildeJet} with the help of the replacement:
\begin{equation} \label{eq:JtildeSign}
    \slashed{p}_i = - \sum_{\sigma_i} v(p_i,-\sigma_i) \bar{u}(p_i,\sigma_i) \; .
\end{equation}

\subsubsection*{Spurious-pole cancellation} \label{sec:poles}

Eq.~\eqref{eq:OneLoopLBK} has been obtained with the expansion-by-regions method. Each region, i.e.\ hard, $(i,j)$-soft and $i$-collinear, contributes {\it spurious poles} in $\epsilon$ due to the unrestricted loop-momentum integration domain. The proof of Eq.~\eqref{eq:OneLoopLBK} is therefore complete when it is shown that all spurious poles cancel. To this end, it is necessary to independently derive an expression for the singularities of the soft-gluon-emission amplitude, expand this result in the soft-gluon momentum and verify agreement with the first two terms of the Laurent expansion of Eq.~\eqref{eq:OneLoopLBK}.

The coefficients of the singular $\epsilon$-expansion terms of an $n$-parton one-loop amplitude $\ket*{M_n^{(1)}(\{k_i\})}$ are contained in the $\mathbf{I}^{(1)}_n$-operator \cite{Giele:1991vf, Kunszt:1994np, Catani:1996jh, Catani:1996vz, Catani:2000ef}:
\begin{equation}
    \ket{M_n^{(1)}(\{k_i\})} = \mathbf{I}^{(1)}_n(\{k_i\}) \ket{M_n^{(0)}(\{k_i\})} + \order{\epsilon^0} \; .
\end{equation}
In the purely massless case, the operator reads:
\begin{equation}
    \mathbf{I}^{(1)}_n(\{k_i\}) = - \frac{1}{\epsilon^2} \sum_i C_i + \frac{1}{\epsilon} \sum_{i \neq j} \mathbf{T}_i \cdot \mathbf{T}_j \ln(- \frac{\mu^2}{2 \, k_i \cdot k_j + i0^+}) + \frac{1}{2 \epsilon} \sum_i \gamma_0^i + \frac{n-2}{2} \frac{\beta_0}{\epsilon} \; .
\end{equation}
The last term proportional to the $\beta$-function coefficient $\beta_0$ is of ultraviolet origin, while the remaining terms are due to soft and collinear singularities. $C_i$ is either the quadratic Casimir operator of the fundamental representation, $C_F = T_F (N_c^2-1) / N_c$, $N_c = 3$, if $i$ is a (anti)-quark, or of the adjoint representation, $C_A = 2 T_F N_c$, if $i$ is a gluon. The anomalous dimensions are given by:
\begin{equation}
    \gamma_0^q = -3 C_F \; ,\qquad \gamma_0^g = - \beta_0 = - \frac{11}{3} C_A + \frac{4}{3} T_F n_l \; .
\end{equation}
For the setup relevant to the present publication, the pole structure reads:
\begin{equation}
\begin{split}
    \ket{M_g^{(1)}(\{p_i+\delta_i\},q)} &= \mathbf{I}^{(1)}_{n+1}\big(\{p_i+\delta_i\},q\big) \ket{M_g^{(0)}(\{p_i+\delta_i\},q)} + \order{\epsilon^0} \\[.2cm]
    &= \mathbf{I}^{(1)}_{n+1}\big(\{p_i+\delta_i\},q\big) \, \Big( \mathbf{S}^{(0)}(\{p_i\},\{\delta_i\},q) \ket{M^{(0)}(\{p_i\})} + \order{\lambda} \Big) + \order{\epsilon^0} \; ,
\end{split}
\end{equation}
with:
\begin{multline}
    \mathbf{P}_g(\sigma,c) \, \mathbf{I}^{(1)}_{n+1}(\{p_i+\delta_i\},q) \, \mathbf{S}^{(0)} \ket{M^{(0)}} = \mathbf{P}_g(\sigma,c) \, \mathbf{S}^{(0)} \, \mathbf{I}_n^{(1)}(\{p_i\}) \ket{M^{(0)}} \\[.2cm]
    + \sum_j \Bigg( \mathbf{T}_j^c \otimes \mathbf{S}^{(0)}_j \, \mathbf{I}_n^{(1)}(\{p_i\}) - \mathbf{I}_n^{(1)}(\{p_i+\delta_i\}) \, \mathbf{T}_j^c \otimes \mathbf{S}^{(0)}_j \\[.2cm]
    + \Bigg( \frac{1}{\epsilon^2} C_A \delta^{cb} - \frac{2}{\epsilon} \sum_i i f^{abc} \mathbf{T}_i^a \ln(- \frac{\mu^2}{s^{(\delta)}_{iq}}) \Bigg) \mathbf{T}_j^b \otimes \mathbf{S}^{(0)}_j \Bigg) \ket{M^{(0)}} \; .
\end{multline}
The r.h.s.\ has already been arranged to exhibit the singularities of the first term in Eq.~\eqref{eq:OneLoopLBK}:
\begin{equation} \label{eq:HardPoles}
    \mathbf{S}^{(0)} \, \ket{M^{(1)}} = \mathbf{S}^{(0)} \, \mathbf{I}_n^{(1)} \ket{M^{(0)}} + \order{\epsilon^0} \; .
\end{equation}
Moreover, we have only made explicit those arguments of the occurring operators that require careful consideration. Further manipulation yields:
\begin{equation} \label{eq:SoftApproxPoles}
\begin{split}
    \mathbf{P}_g(\sigma,c) \, \mathbf{I}^{(1)}_{n+1} \, \mathbf{S}^{(0)} \ket{M^{(0)}} &= \mathbf{P}_g(\sigma,c) \, \mathbf{S}^{(0)} \, \mathbf{I}_n^{(1)} \ket{M^{(0)}} \\[.2cm]
    &+ \frac{2}{\epsilon^2} \sum_{i \neq j} i f^{abc} \mathbf{T}_i^a \mathbf{T}_j^b \otimes \Bigg( 1 + \epsilon \ln( - \frac{\mu^2 s_{ij}^{(\delta)}}{s_{iq}^{(\delta)} s_{jq}^{(\delta)}} ) \Bigg) \mathbf{S}^{(0)}_i \ket{M^{(0)}} \\[.2cm]
    &- \frac{2}{\epsilon} \sum_{i \neq j} \mathbf{T}_i^c \, \mathbf{T}_i \cdot \mathbf{T}_j \, \frac{p_i^\mu p_j^\nu}{p_i \cdot p_j} \frac{iF_{\mu\nu}}{p_i \cdot q} \ket{M^{(0)}} + \order{\lambda} \; .
\end{split}
\end{equation}

Contrary to Eq.~\eqref{eq:OneLoopLBK}, Eq.~\eqref{eq:SoftApproxPoles} does not contain flavour-off-diagonal contributions. Hence, their poles are entirely spurious. We begin the verification of spurious-pole cancellation with the flavour-diagonal contributions.

Expansion of the soft operator \eqref{eq:Soft} acting on the hard matrix element yields:
\begin{multline} \label{eq:SoftPoles}
    \mathbf{P}_g(\sigma,c) \, \mathbf{S}^{(1)} \, \ket{M^{(0)}} = \frac{2}{\epsilon^2} \sum_{i \neq j} i f^{abc} \mathbf{T}_i^a \mathbf{T}_j^b \otimes \Bigg( 1 + \epsilon \ln( - \frac{\mu^2 s_{ij}^{(\delta)}}{s_{iq}^{(\delta)} s_{jq}^{(\delta)}} ) \Bigg) \mathbf{S}^{(0)}_i \ket{M^{(0)}} \\[.2cm]
    + \frac{2}{\epsilon} \, \sum_{i \neq j} i f^{abc} \mathbf{T}_i^a \mathbf{T}_j^b \otimes \frac{1}{p_i \cdot p_j} \Bigg( \frac{p_i^\mu p_j^\nu - p_j^\mu p_i^\nu}{p_i \cdot q} + \frac{p_j^\mu p_j^\nu}{p_j \cdot q} \Bigg) F_{\mu\rho} \, \big( J_i - \mathbf{K}_i \big)^{\nu\rho} \, \ket{M^{(0)}} + \order{\epsilon^0} \; .
\end{multline}
Part of the flavour-diagonal pole contributions generated by the convolution of the jet operator \eqref{eq:Jet} with the collinear-gluon amplitude \eqref{eq:HgDecomposition} is obtained using Eqs.~\eqref{eq:Sgi} and \eqref{eq:Cgi}:
\begin{equation} \label{eq:JetPoles1}
\begin{split}
    \mathbf{P}_g(\sigma,c) \, &\int_0^1 \dd{x} \sum_i \mathbf{J}_i^{(1)} \bigg( \bigg( \frac{1}{x} + \dim(a_i) \bigg) \ket{S^{(0)}_{g,i}} + \ket{C^{(0)}_{g,i}} \bigg) = - \frac{2}{\epsilon} \sum_{i \neq j} \mathbf{T}_i^c \, \mathbf{T}_i \cdot \mathbf{T}_j \, \frac{p_i^\mu p_j^\nu}{p_i \cdot p_j} \frac{iF_{\mu\nu}}{p_i \cdot q} \ket{M^{(0)}} \\[.2cm]
    &- \frac{2}{\epsilon} \, \sum_{i \neq j} i f^{abc} \mathbf{T}_i^a \mathbf{T}_j^b \otimes \frac{1}{p_i \cdot p_j} \Bigg( \frac{p_i^\mu p_j^\nu - p_j^\mu p_i^\nu}{p_i \cdot q} + \frac{p_j^\mu p_j^\nu}{p_j \cdot q} \Bigg) F_{\mu\rho} \, \big( J_i - \mathbf{K}_i \big)^{\nu\rho} \, \ket{M^{(0)}} \\[.2cm]
    &+ \frac{1}{\epsilon} \sum_{i \neq j} \big( 1 - 2\dim(a_i) \big) \, i f^{abc} \mathbf{T}_i^a \mathbf{T}_j^b \otimes \frac{p_i^\rho \, iF_{\rho\mu}}{p_i \cdot q} \bigg( \frac{p_j^\mu}{p_j \cdot p_i} + \bigg( \frac{p_j}{p_j \cdot p_i} - \frac{q}{q \cdot p_i} \bigg)_\sigma \, i \mathbf{K}_i^{\sigma\mu} \bigg) \ket{M^{(0)}} \\[.2cm]
    &+ \order{\epsilon^0} \; .
\end{split}
\end{equation}
If parton $i$ is a gluon, then the soft singularity of the collinear-gluon amplitude at $x = 1$ yields the remaining flavour-diagonal pole contributions. The result is conveniently obtained by rewriting Eq.~\eqref{eq:Jet} in an equivalent form using the Jacobi identity to transform the colour operators and the last of Eqs.~\eqref{eq:Kdef} to transform the spin operator:
%
% Keep original with explicit indices
%
%\begin{equation}
%\begin{split}
%    \mathbf{J}_i^{(1)}(x,p_i,q) & \ket{\dots,c_i',\dots,c';\dots,\sigma_i',\dots,\sigma'} = \\[.2cm]
%    & \frac{\Gamma(1+\epsilon)}{1-\epsilon} \bigg( - \frac{\mu^2}{s_{iq}} \bigg)^{\epsilon} \big( x(1-x) \big)^{-\epsilon} \sum_{cc_i} \bigg( \frac{1-x}{x} \, T^c_{g} T^{c_i'}_{g} + \frac{1}{x} i f^{cdc_i'} T^d_{g} \bigg)_{c_ic'} \\[.2cm]
%    &\times \sum_{\sigma\sigma_i} \epsilon^*_\mu(q,p_i,\sigma) \Big( - 2 (1 - x) \, \epsilon^\mu(p_i,\sigma') \, \delta_{\sigma_i\sigma_i'} - x \, \epsilon^\mu(p_i,\sigma_i') \, \delta_{\sigma_i\sigma'} + x \, \epsilon_\nu(p_i,\sigma_i') \, iK_{g,\sigma_i\sigma'}^{\mu\nu} \Big) \\[.2cm]
%    &\times \ket{\dots,c_i,\dots,c;\dots,\sigma_i,\dots,\sigma} \; .
%\end{split}
%\end{equation}
%
\begin{multline}
    \mathbf{P}_g(\sigma,c) \, \mathbf{J}_i^{(1)}(x,p_i,q) = \frac{\Gamma(1+\epsilon)}{1-\epsilon} \bigg( -\frac{\mu^2}{s_{iq}} \bigg)^{\epsilon} \big( x(1-x) \big)^{-\epsilon} \sum_{\sigma'c'} \epsilon^*_\mu(q,p_i,\sigma) \epsilon_\nu(p_i,\sigma') \\[.2cm]
    \times \Big[ i f^{cdc'} \mathbf{T}^d_i \otimes \big( - g^{\mu\nu} + i \mathbf{K}_i^{\mu\nu} \big) \Big] \, \mathbf{P}_g(\sigma',c') \, \mathbf{E}_{i,n+1} + \text{terms proportional to $(1-x)$} \; .
\end{multline}
Convolution using Eq.~\eqref{eq:Sbargi} yields:
\begin{multline} \label{eq:JetPoles2}
    \mathbf{P}_g(\sigma,c) \, \int_0^1 \dd{x} \sum_i \big( 1 - 2\dim(a_i) \big) \, \mathbf{J}_i^{(1)} \, \frac{x}{1-x} \ket{\bar{S}^{(0)}_{g,i}} = - \frac{1}{\epsilon} \sum_{i \neq j} \big( 1 - 2\dim(a_i) \big) \\[.2cm]
    \times i f^{abc} \mathbf{T}_i^a \mathbf{T}_j^b \otimes \frac{p_i^\rho \, iF_{\rho\mu}}{p_i \cdot q} \bigg( \frac{p_j^\mu}{p_j \cdot p_i} + \bigg( \frac{p_j}{p_j \cdot p_i} - \frac{q}{q \cdot p_i} \bigg)_\sigma i \mathbf{K}_i^{\sigma\mu} \bigg) \ket{M^{(0)}} + \order{\epsilon^0} \; .
\end{multline}
Clearly, the sum of the r.h.s.\ of Eqs.~\eqref{eq:HardPoles}, \eqref{eq:SoftPoles}, \eqref{eq:JetPoles1} and \eqref{eq:JetPoles2} is equal to the r.h.s.\ of Eq.~\eqref{eq:SoftApproxPoles} up to terms of $\order{\lambda}$ and $\order{\epsilon^0}$. This completes the proof of Eq.~\eqref{eq:OneLoopLBK} for the flavour-diagonal contributions.

Let us turn to the poles of flavour-off-diagonal contributions in Eq.~\eqref{eq:OneLoopLBK}, and prove that the poles generated by the flavour-off-diagonal soft operator \eqref{eq:SoftTilde} are cancelled by the poles generated by the convolution of the jet operator \eqref{eq:Jet} with the anti-soft-pole contribution \eqref{eq:Sbargi} for $a_i \in \{q,\bar{q}\}$ and by the convolutions of the flavour-off-diagonal jet operator \eqref{eq:JetTilde} with the soft-pole and anti-soft-pole contributions \eqref{eq:Sqi} and \eqref{eq:Sbarqi}. These three convolutions are given by:
\begin{multline} \label{eq:JetPoles3}
    \int_0^1 \frac{x \dd{x}}{1-x} \, \mel{\dots,c_i,\dots,c;\dots,\sigma_i,\dots,\sigma}{\mathbf{J}_i^{(1)}}{\bar{S}^{(0)}_{g,i}} = \frac{1}{\epsilon} \, \epsilon^*_\mu(q,p_i,\sigma) \sum_{j \neq i} \sum_{\sigma_i'c_i'} \sum_{\sigma_j'c_j'} \sum_{\sigma'c'} \\[.2cm] \epsilon_\nu(p_i,\sigma_i') \big( T_{a_i}^{c_i'} T_{a_i}^c \big)_{c_i c'} \big( g^{\mu\nu} \mathbbm{1} - 2iK_{a_i}^{\mu\nu}(p_i) \big)_{\sigma_i\sigma'} \mel{c_j,c';\sigma_j,\sigma'}{\mathbf{Split}^{(0)}_{a_ja_i \leftarrow \tilde{a}_j}(p_j,p_i,p_j)}{c_j';\sigma_j'} \\[.2cm] \times \ip{\dots,c_i',\dots,c_j',\dots;\dots,\sigma_i',\dots,\sigma_j',\dots}{M^{(0)}(\{p_i\}) \, \Big|\substack{a_i \, \to \, g \;\; \\ a_j \, \to \, \tilde{a}_j}} + \order{\epsilon^0} \\[.2cm]
    \equiv \frac{1}{\epsilon} \sum_{j \neq i} \sum_{\sigma_i'c_i'} \sum_{\sigma_j'c_j'} J^{(1,-1)}_{a_i a_j \, \leftarrow \, g \tilde{a}_j} \ip{\dots,c_i',\dots,c_j',\dots;\dots,\sigma_i',\dots,\sigma_j',\dots}{M^{(0)}(\{p_i\}) \, \Big|\substack{a_i \, \to \, g \;\; \\ a_j \, \to \, \tilde{a}_j}} + \order{\epsilon^0} \; ,
\end{multline}
\begin{multline} \label{eq:JetPoles4}
    \int_0^1 \dd{x} \, \frac{1}{x} \, \mel{\dots,c_i,\dots,c;\dots,\sigma_i,\dots,\sigma}{\mathbf{\tilde{J}}_i^{(1)}}{S^{(0)}_{\bar{q},i}} = \frac{1}{\epsilon} \, \epsilon^*_\mu(q,p_i,\sigma) \epsilon^*_\nu(p_i,\sigma_i) \sum_{j \neq i} \sum_{\sigma_i'c_i'} \sum_{\sigma_j'c_j'} \sum_{\sigma'c'} \\[.2cm] \big( T^c T^{c_i} \big)_{c'c_i'} \big( - g^{\mu\nu} \, \mathbbm{1} - 2 \, iK_q^{\mu\nu}(p_i) \big)_{-\sigma'\sigma_i'} \mel{c_j,c';\sigma_j,\sigma'}{\mathbf{Split}^{(0)}_{a_j\bar{q} \, \leftarrow \, \tilde{a}_j}(p_j,p_i,p_j)}{c_j';\sigma_j'} \\[.2cm] \times \ip{\dots,c_i',\dots,c_j',\dots;\dots,\sigma_i',\dots,\sigma_j',\dots}{M^{(0)}(\{p_i\}) \, \Big|\substack{a_i \, \to \, q \;\; \\ a_j \, \to \, \tilde{a}_j}} + \order{\epsilon^0} \\[.2cm]
    \equiv \frac{1}{\epsilon} \sum_{j \neq i} \sum_{\sigma_i'c_i'} \sum_{\sigma_j'c_j'} \tilde{J}^{(1,-1)}_{a_i a_j \, \leftarrow \, q \tilde{a}_j} \ip{\dots,c_i',\dots,c_j',\dots;\dots,\sigma_i',\dots,\sigma_j',\dots}{M^{(0)}(\{p_i\}) \, \Big|\substack{a_i \, \to \, q \;\; \\ a_j \, \to \, \tilde{a}_j}} + \order{\epsilon^0} \; ,
\end{multline}
\begin{multline} \label{eq:JetPoles5}
    \int_0^1 \dd{x} \, \frac{x}{1-x} \, \mel{\dots,c_i,\dots,c;\dots,\sigma_i,\dots,\sigma}{\mathbf{\tilde{J}}_i^{(1)}}{\bar{S}^{(0)}_{\bar{q},i}} = \frac{1}{\epsilon} \, \epsilon^*_\mu(q,p_i,\sigma) \epsilon^*_\nu(p_i,\sigma_i) \sum_{j \neq i} \sum_{\sigma_i'c_i'} \sum_{\sigma_j'c_j'}\sum_{\sigma'c'} \\[.2cm] \big( T^{c_i} T^c \big)_{c_i'c'} \big( g^{\mu\nu} \, \mathbbm{1} - 2 \, iK_q^{\mu\nu}(p_i) \big)_{-\sigma_i'\sigma'} \mel{c_j,c';\sigma_j,\sigma'}{\mathbf{Split}^{(0)}_{a_jq \, \leftarrow \, \tilde{a}_j}(p_j,p_i,p_j)}{c_j';\sigma_j'} \\[.2cm] \times \ip{\dots,c_i',\dots,c_j',\dots;\dots,\sigma_i',\dots,\sigma_j',\dots}{M^{(0)}(\{p_i\}) \, \Big|\substack{a_i \, \to \, \bar{q} \;\; \\ a_j \, \to \, \tilde{a}_j}} + \order{\epsilon^0} \\[.2cm]
    \equiv \frac{1}{\epsilon} \sum_{j \neq i} \sum_{\sigma_i'c_i'} \sum_{\sigma_j'c_j'} \tilde{J}^{(1,-1)}_{a_i a_j \, \leftarrow \, \bar{q} \tilde{a}_j} \ip{\dots,c_i',\dots,c_j',\dots;\dots,\sigma_i',\dots,\sigma_j',\dots}{M^{(0)}(\{p_i\}) \, \Big|\substack{a_i \, \to \, \bar{q} \;\; \\ a_j \, \to \, \tilde{a}_j}} + \order{\epsilon^0} \; .
\end{multline}
Substitution of the splitting operators listed in Section~\ref{sec:SplittingOperators} and application of the definitions \eqref{eq:Kdef} of the spin operators yields:
\begin{align} \label{eq:JetPoles345}
    J^{(1,-1)}_{q \bar{q} \, \leftarrow \, g g} &= -\frac{1}{2 \, p_i \cdot p_j} \big( T^{c_i'} T^c T^{c_j'} \big)_{c_i c_j} \, \bar{u}(p_i,\sigma_i) \slashed{\epsilon}(p_i,\sigma_i') \slashed{\epsilon}^*(q,p_i,\sigma) \slashed{\epsilon}(p_j,\sigma_j') v(p_j,\sigma_j) \; , \nonumber\\[.2cm]
    J^{(1,-1)}_{q g \, \leftarrow \, g q} &= -\frac{1}{2 \, p_i \cdot p_j} \big( T^{c_i'} T^c T^{c_j} \big)_{c_i c_j'} \, \bar{u}(p_i,\sigma_i) \slashed{\epsilon}(p_i,\sigma_i') \slashed{\epsilon}^*(q,p_i,\sigma) \slashed{\epsilon}^*(p_j,\sigma_j) u(p_j,\sigma_j') \; , \nonumber\\[.2cm]
    J^{(1,-1)}_{\bar{q} g \, \leftarrow \, g \bar{q}} &= + \frac{1}{2 \, p_i \cdot p_j} \big( T^{c_j} T^c T^{c_i'} \big)_{c_j' c_i} \, \bar{v}(p_j,\sigma_j') \slashed{\epsilon}^*(p_j,\sigma_j) \slashed{\epsilon}^*(q,p_i,\sigma) \slashed{\epsilon}(p_i,\sigma_i') v(p_i,\sigma_i) \; , \nonumber\\[.2cm]
    \tilde{J}^{(1,-1)}_{g q \, \leftarrow \, q g} &= + \frac{1}{2 \, p_i \cdot p_j} \big( T^{c_j'} T^c T^{c_i} \big)_{c_j c_i'} \, \bar{u}(p_j,\sigma_j) \slashed{\epsilon}(p_j,\sigma_j') \slashed{\epsilon}^*(q,p_i,\sigma) \slashed{\epsilon}^*(p_i,\sigma_i) v(p_i,-\sigma_i') \; , \\[.2cm]
    \tilde{J}^{(1,-1)}_{g g \, \leftarrow \, q \bar{q}} &= -\frac{1}{2 \, p_i \cdot p_j} \big( T^{c_j} T^c T^{c_i} \big)_{c_j' c_i'} \, \bar{v}(p_j,\sigma_j') \slashed{\epsilon}^*(p_j,\sigma_j) \slashed{\epsilon}^*(q,p_i,\sigma) \slashed{\epsilon}^*(p_i,\sigma_i) v(p_i,-\sigma_i') \; , \nonumber\\[.2cm]
    \tilde{J}^{(1,-1)}_{g g \, \leftarrow \, \bar{q} q} &= -\frac{1}{2 \, p_i \cdot p_j} \big( T^{c_i} T^c T^{c_j} \big)_{c_i' c_j'} \, \bar{u}(p_i,-\sigma_i')\slashed{\epsilon}^*(p_i,\sigma_i) \slashed{\epsilon}^*(q,p_i,\sigma) \slashed{\epsilon}^*(p_j,\sigma_j) u(p_j,\sigma_j') \; , \nonumber\\[.2cm]
    \tilde{J}^{(1,-1)}_{g \bar{q} \, \leftarrow \, \bar{q} g} &= -\frac{1}{2 \, p_i \cdot p_j} \big( T^{c_i} T^c T^{c_j'} \big)_{c_i' c_j} \, \bar{u}(p_i,-\sigma_i') \slashed{\epsilon}^*(p_i,\sigma_i) \slashed{\epsilon}^*(q,p_i,\sigma) \slashed{\epsilon}(p_j,\sigma_j') v(p_j,\sigma_j) \; .\nonumber
\end{align}
Bi-spinors depending on $-\sigma_i'$ are subsequently replaced by bi-spinors depending on $+\sigma_i'$ according to Eq.~\eqref{eq:UVrelation}. The resulting expressions can be further simplified using:
\begin{equation}
\begin{aligned}
    &\dots \slashed{\epsilon}^*(q,p_i,\sigma) \dots = - \frac{1}{2 p_i \cdot p_j} \dots \slashed{p}_j \slashed{\epsilon}^*(q,p_i,\sigma) \slashed{p}_i \dots \qquad \text{or} \\[.2cm]
    &\dots \slashed{\epsilon}^*(q,p_i,\sigma) \dots = - \frac{1}{2 p_i \cdot p_j} \dots \slashed{p}_i \slashed{\epsilon}^*(q,p_i,\sigma) \slashed{p}_j \dots \; ,
\end{aligned}
\end{equation}
where the dots stand for the factors occurring in Eqs.~\eqref{eq:JetPoles345}, and the first equality applies if the left factor depends on $p_i$, while the second equality applies if the left factor depends on $p_j$. It can now be easily verified using:
\begin{equation}
\begin{gathered}
    \sum_{\sigma_i''} v(p_i,\sigma_i'') \bar{v}(p_i,\sigma_i'') = \slashed{p}_i \; , \qquad
    \sum_{\sigma_i''} u(p_i,\sigma_i'') \bar{u}(p_i,\sigma_i'') = \slashed{p}_i \; , \\[.2cm]
    \sum_{\sigma_j''} v(p_j,\sigma_j'') \bar{v}(p_j,\sigma_i'') = \slashed{p}_j \; , \qquad
    \sum_{\sigma_j''} u(p_j,\sigma_j'') \bar{u}(p_j,\sigma_i'') = \slashed{p}_j \; ,
\end{gathered}
\end{equation}
that each pole coefficient listed in \eqref{eq:JetPoles345} cancels a respective pole coefficient in Eq.~\eqref{eq:SoftTilde}. This completes the proof of Eq.~\eqref{eq:OneLoopLBK} for the flavour-off-diagonal contributions.

\subsection{Numerical tests}

\begin{figure}
    \centering
    \includegraphics[width=\textwidth]{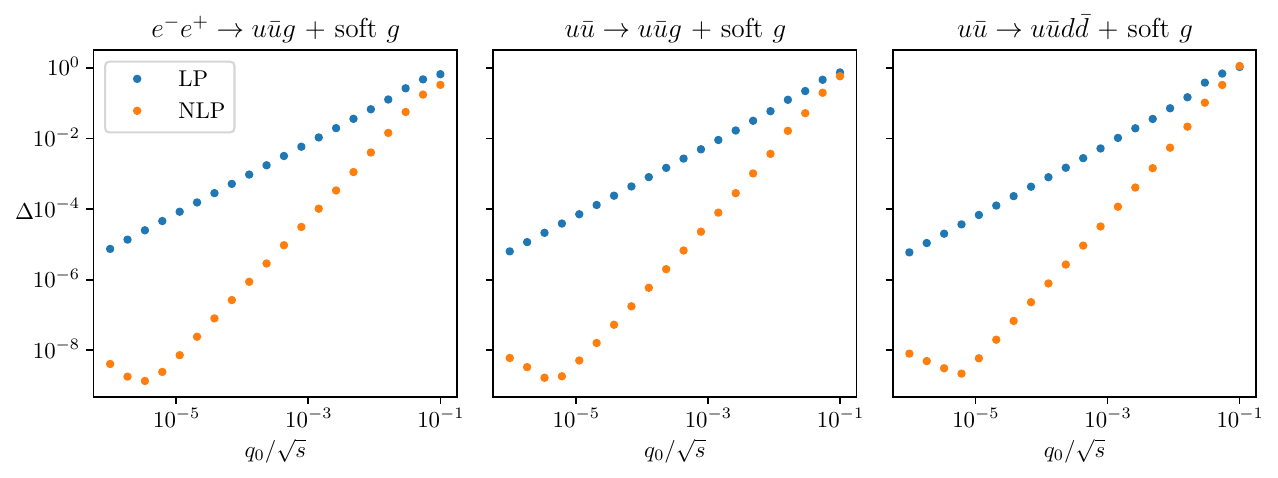}
    \caption{Relative error $\Delta_{\text{LP/NLP}}$ of the one-loop soft approximation to leading power (LP) and subleading power (NLP). The energy, $q_0$, of the soft gluon is normalised to the centre-of-mass energy, $\sqrt{s}$, of the process. The apparent breakdown of the approximation at low soft-gluon energies is due to the limited numerical precision of the one-loop integrals in \textsc{OneLOop} which impacts the result for the $(n+1)$-particle amplitudes.}
    \label{fig:numerics_main}
\end{figure}

\begin{figure}
    \centering
    \includegraphics[width=.66\textwidth]{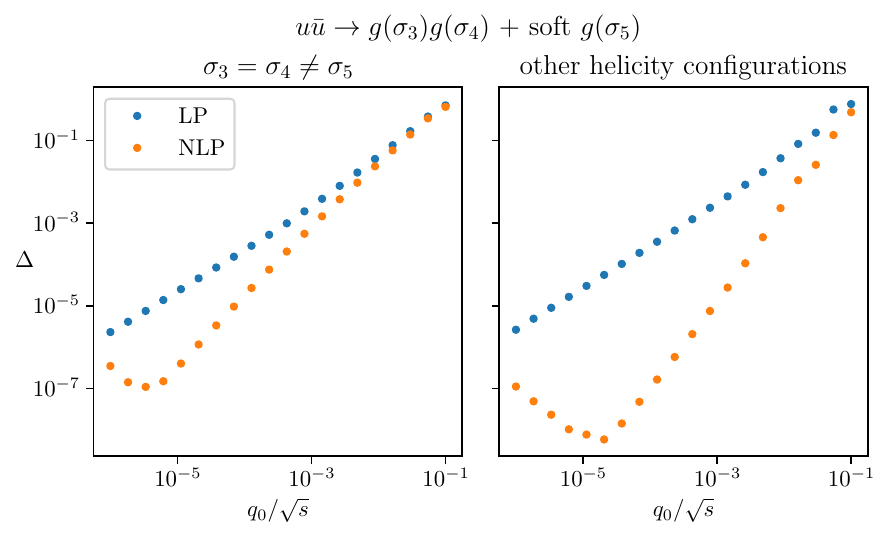}
    \caption{Plots analogous to Fig.~\ref{fig:numerics_main} except that the helicity sum is restricted to a specific setup in the left plot, and the right plot contains all other helicity configurations.}
    \label{fig:numerics_gluons}
\end{figure}

Although theorem \eqref{eq:OneLoopLBK} has been strictly proven in Section~\ref{sec:proof}, it is still a useful and instructive exercise to verify the formulae of Sections~\ref{sec:theorem}, \ref{sec:CollinearAmplitudes} and \ref{sec:convolutions} on actual amplitudes. In this section, we numerically evaluate the $\order{\epsilon^0}$ coefficient of the Laurent expansion of $\ket{M_g^{(1)}}$ for several processes and compare it to the result of the soft expansion. For a stringent test, we consider processes that involve up to six hard partons, both incoming and outgoing, multiple quark flavours and colour-neutral particles. The list can be read off of Figs.~\ref{fig:numerics_main} and \ref{fig:numerics_gluons}.

Let us define the difference between the exact and the approximate amplitude:
\begin{equation} \label{eq:DeltaNLP}
    \Delta_\text{LP/NLP} \equiv \frac{1}{N} \sum_{\substack{\text{singular} \\ \text{colour flows $\{c\}$} \\ \text{helicities $\{\sigma\}$}}} \left|\frac{\left[\braket{\{c,\sigma\}}{M_{g}^{(1)}} - \braket{\{c,\sigma\}}{M^{(1)}_g}_\text{LP/NLP}\right]_{\mathcal{O}(\epsilon^0)}}{\left[\braket{\{c,\sigma\}}{M_{g}^{(1)}}\right]_{\mathcal{O}(\epsilon^0)}}\right| \; ,
\end{equation}
where LP (leading power) stands for soft expansion up to $\order{1/\lambda}$, while NLP (next-to-leading power) up to $\order{\lambda^0}$. The sum runs over all colour-flow and helicity configurations for which the amplitude has a soft singularity. The number of such configurations is denoted by $N$. The one-loop $n$-particle amplitudes $\ket{M^{(1)}}$ as well as their derivatives are calculated with \textsc{Recola}~\cite{Actis:2016mpe, Denner:2017wsf} linked to \textsc{Collier}~\cite{Denner:2016kdg, Denner:2002ii, Denner:2005nn, Denner:2010tr} for the evaluation of tensor and scalar one-loop integrals. For the evaluation of the one-loop $(n+1)$-particle amplitudes, $\ket{M_g^{(1)}}$, we instead link \textsc{Recola} to \textsc{CutTools}~\cite{Ossola:2007ax} for tensor reduction and \textsc{OneLOop}~\cite{vanHameren:2009dr,vanHameren:2010cp} for the evaluation of scalar integrals at quadruple precision. Finally, for the evaluation of the collinear amplitudes, we use Eqs.~\eqref{eq:DeltaMgDirect} and \eqref{eq:HqDirect} implemented by calling \textsc{AvH}~\cite{Bury:2015dla} with replaced spinors and polarisation vectors of the external particles as appropriate. The $x$-dependence of the collinear amplitudes is obtained at first by rational-function fitting. Subsequently, we verify that the results agree with those obtained by direct evaluation with the formulae from the last paragraph of Section~\ref{sec:CollinearAmplitudes}. A subtlety arises from the fact that amplitudes for different processes are involved in the computation of \eqref{eq:DeltaNLP}. Indeed, the global sign of the amplitudes depends on the external fermion ordering and the algorithm used. Therefore, for the flavour-off-diagonal contributions, we have to compensate the differences between the software tools by including appropriate signs to obtain the correct result.


$\Delta_{\text{LP/NLP}}$ is expected to have the following behaviour:
\begin{gather}
    \Delta_\text{LP} = \left(c_0+c_1\log\lambda+c_2\log^2\lambda\right)\lambda + \mathcal{O}(\lambda^2) \; , \\[.2cm]
    \Delta_\text{NLP} = \left(d_0+d_1\log\lambda+d_2\log^2\lambda\right)\lambda^2 + \mathcal{O}(\lambda^3) \; .
    \label{eq:Delta_scaling}
\end{gather}
This behaviour is reproduced for the three example processes in Fig.~\ref{fig:numerics_main} as much as numerical precision permits. Fig.~\ref{fig:numerics_gluons} shows, split by helicity configuration, the results for the process:
\begin{equation}\label{eq:full_process_2g}
    q(\sigma_1)+\bar{q}(\sigma_2)\to g(\sigma_3)+g(\sigma_4)+g(q,\sigma_5),
\end{equation}
where $q$ is the soft momentum, and hard-momentum and colour dependence are suppressed for brevity. For most configurations, the test results show a strong improvement between LP and NLP in line with Fig.~\ref{fig:numerics_main}. However, in the case $\sigma_3=\sigma_4\neq\sigma_5$, the improvement is less pronounced while still remaining consistent with \eqref{eq:Delta_scaling}. This spin configuration is distinguished by the fact that it does not contain any logarithms containing the soft momentum through next-to-leading power. For example, the flavour-diagonal soft-region contribution is proportional to the tree-level amplitude of the process:
\begin{equation}
    q(\sigma_1)+\bar{q}(\sigma_2)\to g(\sigma_3)+g(\sigma_4),
\end{equation}
which vanishes if $\sigma_3=\sigma_4$ due to helicity conservation. It is not hard to convince oneself that all flavour-off-diagonal soft-region contributions vanish in full analogy. The flavour-diagonal collinear region does not contribute because the collinear hard function is derived from the subleading collinear behaviour of the process:
\begin{equation}
    q(\sigma_1)+\bar{q}(\sigma_2)\to g(\sigma_3)+g(\sigma_4)+g(-\sigma_5),
\end{equation}
which follows from the full process definition \eqref{eq:full_process_2g} and the properties of the jet operator. In particular, the occurrence of $-\sigma_5$ can be conveniently read off Eq.~\eqref{eq:JetConvolution}. Again, this process vanishes at tree level for $\sigma_3=\sigma_4\neq\sigma_5$ due to helicity conservation. Finally, the flavour-off-diagonal jet operator is only non-zero if $\sigma_i=\sigma_5$ for $a_i=g$, i.e.\ $i\in\{3,4\}$, which is not fulfilled for the considered helicity configuration. Altogether, only the hard-region contribution to Eq.~\eqref{eq:OneLoopLBK}, $\mathbf{S}^{(0)}\ket{M^{(1)}}$, is non-zero for the considered spin configuration. While next-to-next-to-leading-power contributions to the soft expansion are not discussed in the present publication, the behaviour observed in Fig.~\ref{fig:numerics_gluons} shows that one can expect soft logarithms starting to appear there, implying a less-constrained helicity structure. The poorer numerical behaviour is not expected to pose a problem in practical applications because for squared amplitudes summed over colour and helicity, the helicity configurations which contain soft logarithms already at leading power dominate numerically in the soft momentum region.

\section{Next-to-leading-power collinear asymptotics at tree-level} \label{sec:CollinearNLP}

The collinear-gluon and collinear-quark amplitude constructed in Section~\ref{sec:CollinearAmplitudes} may be used to derive a result for the collinear asymptotics of massless tree-level QCD amplitudes that correctly accounts for subleading effects. We consider the collinear limit for partons $i$ and $n+1$:
\begin{align}
    &k_{n+1} \equiv x p_i + l_\perp - \frac{l_\perp^2}{2x} \frac{q}{p_i \cdot q} \; , & \text{with} & \qquad l_\perp \cdot p_i = l_\perp \cdot q = 0 \; , \\[.2cm]
    & k_i \equiv (1-x) p_i - l_\perp - \frac{l_\perp^2}{2(1-x)} \frac{q}{p_i \cdot q} \; , & \text{and}
    & \qquad k_j \equiv p_j + \order{l_\perp^2} \; , \qquad j \neq i \; .
\end{align}
For $a_i = a_{n+1} = g$, the collinear expansion is given by:
\begin{align}
    \mathbf{P}_i(\sigma_i,c_i) & \mathbf{P}_{n+1}(\sigma_{n+1},c_{n+1}) \ket{M^{(0)}(\{k_i\}_{i=1}^{n+1})} = 
    \mathbf{P}_i(\sigma_i,c_i) \mathbf{P}_{n+1}(\sigma_{n+1},c_{n+1}) \bigg[ \nonumber\\[.2cm]
    &\mathbf{Split}^{(0)}_{i,n+1 \, \leftarrow \, i}(k_i,k_{n+1},p_i) \ket{M^{(0)}(\{p_i\})} \nonumber\\[.2cm]
    &+ \bigg( \frac{1-x^2}{x} + \frac{1-(1-x)^2}{1-x} \mathbf{E}_{i,n+1} \bigg) \, \ket{S^{(0)}_{g,i}(\{p_i\},q)} + \big( (1-x) + x \mathbf{E}_{i,n+1} \big) \ket{C^{(0)}_{g,i}(\{p_i\},q)} \nonumber\\[.2cm]
    &+ \frac{1}{2} \sum_I \frac{x(1-x)}{x_I(1-x_I)} \bigg( \frac{1}{x_I - x} + \frac{1}{x_I - (1-x)} \mathbf{E}_{i,n+1} \bigg) \ket{R^{(0)}_{g,i,I}(\{p_i\})} \bigg] \\[.2cm]
    &+ \bigg[ \frac{1}{x} \frac{q \cdot \epsilon^*(p_i,\sigma_{n+1})}{q \cdot p_i} \mathbf{P}_i(\sigma_i,c_i) \mathbf{T}_i^{c_{n+1}} + \frac{1}{1-x} \frac{q \cdot \epsilon^*(p_i,\sigma_i)}{q \cdot p_i} \mathbf{P}_i(\sigma_{n+1},c_{n+1}) \mathbf{T}_i^{c_i} \bigg] \ket{M^{(0)}(\{p_i\})} \nonumber\\[.2cm]
    &+ \order{l_\perp} \; ,\nonumber
\end{align}
with $\ket*{S^{(0)}_{g,i}(\{p_i\},q)}$, $\ket*{C^{(0)}_{g,i}(\{p_i\},q)}$ and $\ket*{R^{(0)}_{g,i,I}(\{p_i\})}$ defined in Eqs.~\eqref{eq:Sgi}, \eqref{eq:Cgi} and \eqref{eq:RgiI} respectively. The splitting function acting on $\ket*{M^{(0)}(\{p_i\})}$ introduces a helicity sum for the intermediate gluon. This sum must be consistent with Eq.~\eqref{eq:HelSumSplit}. We note that the subleading collinear asymptotics requires the subleading soft asymptotics contained in $\ket*{C^{(0)}_{g,i}(\{p_i\},q)}$.

\noindent
For $a_i \in \{q,\bar{q}\}$, $a_{n+1} = g$, one finds:
\begin{equation}
\begin{split}
    \mathbf{P}_{n+1}&(\sigma_{n+1},c_{n+1}) \ket{M^{(0)}(\{k_i\}_{i=1}^{n+1})} = 
    \mathbf{P}_{n+1}(\sigma_{n+1},c_{n+1}) \bigg[ \\[.2cm]
    &\mathbf{Split}^{(0)}_{i,n+1 \, \leftarrow \, i}(k_i,k_{n+1},p_i) \ket{M^{(0)}(\{p_i\})} \\[.2cm]
    &+ \sqrt{1-x} \bigg( \bigg( \frac{1}{x} + \frac{1}{2} \bigg) \, \ket{S^{(0)}_{g,i}(\{p_i\},q)} + \ket{C^{(0)}_{g,i}(\{p_i\},q)} + \frac{x}{1-x} \ket{\bar{S}^{(0)}_{g,i}(\{p_i\},q)} \\[.2cm]
    &+ \sum_I \bigg( \frac{1}{x_I - x} - \frac{1}{x_I} \bigg) \ket{R^{(0)}_{g,i,I}(\{p_i\})} \bigg) \bigg]
    + \frac{\sqrt{1-x}}{x} \frac{q \cdot \epsilon^*(p_i,\sigma_{n+1})}{q \cdot p_i} \mathbf{T}_i^{c_{n+1}} \ket{M^{(0)}(\{p_i\})} \\[.2cm]
    &+ \order{l_\perp} \; .
\end{split}
\end{equation}

\noindent
Finally, for $a_i =q$, $a_{n+1} = \bar{q}$, one finds:
\begin{multline}
    \ket{M^{(0)}(\{k_i\}_{i=1}^{n+1})} = 
    \mathbf{Split}^{(0)}_{i,n+1 \, \leftarrow \, i}(k_i,k_{n+1},p_i) \ket{M^{(0)}(\{p_i\})} \\[.2cm]
    + \sqrt{x(1-x)} \bigg( \frac{1}{x} \ket{S^{(0)}_{\bar{q},i}(\{p_i\})} + \ket{C^{(0)}_{\bar{q},i}(\{p_i\},q)} + \frac{x}{1-x} \ket{\bar{S}^{(0)}_{\bar{q},i}(\{p_i\})} \\[.2cm]
    + \sum_I \bigg( \frac{1}{x_I - x} - \frac{1}{x_I} \bigg) \ket{R^{(0)}_{\bar{q},i,I}(\{p_i\})} \bigg) + \order{l_\perp} \; .
\end{multline}
Since the splitting proceeds via an intermediate gluon, the occurring helicity sum must be consistent with Eq.~\eqref{eq:HelSumSplit}. The contributions $\ket*{S^{(0)}_{\bar{q},i}(\{p_i\})}$, $\ket*{\bar{S}^{(0)}_{\bar{q},i}(\{p_i\})}$ and $\ket*{R^{(0)}_{\bar{q},i,I}(\{p_i\})}$ are defined in Eqs.~\eqref{eq:Sqi}, \eqref{eq:Sbarqi} and \eqref{eq:RqiI} respectively. As remarked at the end of Section~\ref{sec:CollinearAmplitudes}, the contribution $\ket*{C^{(0)}_{\bar{q},i}(\{p_i\},q)}$ corresponds to the subleading term of the soft-anti-quark expansion of the collinear-quark amplitude. As we do not provide an explicit expression in terms of $\ket*{M^{(0)}(\{p_i\})}$ for this contribution, it must be evaluated by using Eq.~\eqref{eq:HqDirect} at a convenient point in $x$.

\section{Summary and outlook}

This publication contains two novel results. The first one is the general formula for the approximation of a one-loop soft-gluon emission amplitude at next-to-leading power presented in Section~\ref{sec:OneLoopLBK}. The second are the general formulae for the approximation of tree-level amplitudes in the collinear limit at next-to-leading power presented in Section~\ref{sec:CollinearNLP}. Both results are limited to massless partons, but allow for the inclusion of arbitrary colour-neutral particles. They are expressed through universal factors and process-dependent gauge-invariant amplitudes. As such, they cannot be further simplified.

It is interesting to note that the tree-level collinear approximations require the knowledge of the tree-level soft approximations, while the one-loop soft approximation requires the knowledge of both the tree-level collinear and soft approximation. We expect this pattern to extend to higher orders, i.e.\ higher order soft approximations should depend on lower order collinear approximations. In any case, extension of the results to higher orders is one natural direction for future research.

We must point out once more that the provided next-to-leading power approximation for a collinear quark-anti-quark pair requires the subleading soft term of the soft-anti-quark expansion of the collinear amplitude, for which no general formula is known at present. In practice, one can obtain the necessary result by a single evaluation of a suitably prepared amplitude at fixed kinematics. Nevertheless, it would be much more elegant to have an expression similar to the LBK theorem. We leave this problem to future work.

Our results should be extended to massive partons in a next step. On the one hand, this extension should be simpler by not containing collinear regions and flavour-off-diagonal contributions for massive partons. On the other hand, the difference between the leading soft asymptotics for massless \cite{Catani:2000pi} and massive partons \cite{Bierenbaum:2011gg, Czakon:2018iev} suggests that the expression for the soft operator will be much more complex in the massive case.

\begin{acknowledgments}
We would like to thank Daniel Stremmer for help with linking \textsc{Recola} to \textsc{CutTools} and \textsc{OneLOop}. This work was supported by the Deutsche Forschungsgemeinschaft (DFG) under grant 396021762 - TRR 257: Particle Physics Phenomenology after the Higgs Discovery, and grant 400140256 - GRK 2497: The Physics of the Heaviest Particles at the LHC. Diagrams were drawn using \textsc{JaxoDraw}~\cite{Vermaseren:1994je, Collins:2016aya,Binosi:2003yf}.
\end{acknowledgments}

\newpage
\bibliographystyle{JHEP}
\bibliography{main} 

\end{document}